\RequirePackage{rotating} 
\documentclass[acmsmall,screen,authorversion,nonacm]{acmart}

\AtBeginDocument{%
  \providecommand\BibTeX{{%
    \normalfont B\kern-0.5em{\scshape i\kern-0.25em b}\kern-0.8em\TeX}}}

\setcopyright{acmcopyright}
\copyrightyear{2021}
\acmYear{2021}
\acmDOI{10.1145/1122445.1122456}

\newcommand{\LinGBM}
	{LinGBM}

\newcommand{\todo}[1]{}

\usepackage{multirow}
\usepackage{fnpct}
\usepackage{hhline}
\usepackage{listings}
\usepackage{rotating}
\usepackage{subfig}

\newcommand{\tnote}[1]{$^\text{#1}$}

\definecolor{darkgray}{rgb}{.3,.3,.3}

\newcommand{\FinalVersion}[1]{#1}\newcommand{\DraftVersion}[1]{}

\FinalVersion{}
\DraftVersion{}

\FinalVersion{\newcommand{\removable}[1]{#1}}
\DraftVersion{\newcommand{\removable}[1]{{\color{darkgray}#1}}}

\newcommand{\hidden}[1]{}

\newcommand{\removableAltOff}[2]{\removable{#1}}  

\newcommand{\sfactor}{\mathit{sf}}

\newcommand{\pholder}[1]%
	{\textit{#1}}

\newcommand{\metric}[1]{\textsf{\small #1}}  
\newcommand{\aQETq}{\metric{aQETq}}
\newcommand{\aQRTq}{\metric{aQRTq}}
\newcommand{\QETt}{\metric{QETt}}
\newcommand{\aTPt}{\metric{aTPt}}
\newcommand{\aTPw}{\metric{aTPw}}
\newcommand{\aTPm}{\metric{aTPm}}

\begin{document}

\title{\LinGBM: A Performance Benchmark for Approaches to Build GraphQL Servers (Extended Version)}

\titlenote{This article is an extended version of a research paper with the same title, published in the proceedings of the 23rd~International Conference on Web Information Systems Engineering~(WISE~2022). The extension consists of a background section that provides an overview of GraphQL~(cf.\ Section~\ref{ssec:Background:OverviewOfGraphQL}) and of approaches to create GraphQL servers~(cf.\ Section~\ref{ssec:Background:Approaches}), an overview of the tools that we have developed to enable users to perform experiments with the benchmark~(cf.\ Section~\ref{ssec:Artifacts:Tools}), a more detailed discussion of statistical properties of the benchmark datasets and queries~(cf.\ Section~\ref{sec:Properties}), and an additional use case in which we have applied the benchmark~(cf.\ Section~\ref{subsec:UC3}).}

\author{Sijin Cheng}
\email{sijin.cheng@liu.se}
\orcid{0003-4363-0654}
\affiliation{%
  \institution{Link\"oping University}
  \city{Link\"oping}
  \country{Sweden}
}

\author{Olaf Hartig}
\email{olaf.hartig@liu.se}
\orcid{0000-0002-1741-2090}
\affiliation{%
  \institution{Link\"oping University}
  \city{Link\"oping}
  \country{Sweden}
}

\begin{abstract}
	GraphQL is a popular new approach to build Web APIs that enable clients to retrieve exactly the data they need. Given
the growing number of tools and techniques for building GraphQL servers, there is an increasing need for comparing how particular approaches or techniques affect the performance of a GraphQL server. To this end, we present \LinGBM, a GraphQL performance benchmark to experimentally study the performance achieved by various approaches for creating a GraphQL server. In this article, we
	discuss the design considerations
of the benchmark, describe
	its main components
%
(data schema; query templates; performance metrics), and
	analyze the benchmark in terms of statistical properties that are relevant for defining concrete experiments.
Thereafter, we present experimental results obtained by applying the benchmark in three different use cases, which demonstrates the
	broad applicability of \LinGBM.
\end{abstract}



\maketitle

\section{Introduction} \label{sec:Introduction}

GraphQL is a new approach to build
	data access APIs for Web and mobile applications%
~\cite{GraphQLSpec}.
Since
	its first published
specification in July~2015, the approach has become tremendously popular with a flourishing ecosystem of related programming libraries and software tools~\cite{GraphQLlandscape}, and many adopters. For instance,
	a study of GraphQL in
open source projects identified more than 37,000~code repositories that depend on the GraphQL reference implementation~\cite{DBLP:conf/amw/KimCH19}, which is just one of several implementations of the approach.
	A
similar study found 8,399 unique GraphQL API schemas on Github~\cite{DBLP:conf/icsoc/WitternCDBM19}. Besides open source projects,
	many companies are adopting GraphQL\removable{ for their commercial software applications}~\cite{GraphQLlandscape}. In 2019, some of these adopters have formed the GraphQL Foundation to financially support future standardization and development of the GraphQL approach, \-including household names such as Airbnb, AWS, Expedia, Facebook, Goldman Sachs, IBM, Paypal, Shopify, and Twitter~\cite{GraphQLFoundationAnnualReport}.

What makes
	GraphQL
interesting from a
	systems
research perspective is that it is based on a declarative query language which enables clients to define precisely the data they want to retrieve.
	The advantage of this approach, in comparison to REST\hidden{-based} interfaces\removable{~\cite{richardson2013restful}}, is that it 
reduces both the number of requests that need to be issued by clients and the amount of data transferred between server and client%
	~\cite{DBLP:conf/wcre/BritoMV19,DBLP:conf/icsa/BritoV20}.

To leverage this advantage, however, it requires GraphQL servers that can process
	the given
query requests~efficiently.
%
There exists a plethora of Web tutorials and blog posts, as well as several books~(e.g., \cite{Porcello:GraphQLbook, Grebe:GraphQLbook, Buna:GraphQLbook1, Buna:GraphQLbook2, Kimokoti:GraphQLbook, Williams:GraphQLbook}), that all describe approaches to implement a GraphQL server, including techniques to avoid typical performance pitfalls and to optimize various aspects of the implemented server.
	However,
studies that show
	\removable{or even compare} how using particular approaches or techniques
may affect the performance of the resulting GraphQL server are rare and remain often anecdotal. However, understanding \hidden{this impact and }the pros and cons of different solutions is crucial \hidden{to choose the right options }for building an \emph{efficient} GraphQL server that
	provides
an optimal performance for a given~application.

Achieving such an understanding requires
	performance tests,
for which suitable experimentation frameworks, methods, and tooling are needed. While there
	are
a few per\-for\-mance-re\-lat\-ed
	test suites for specific GraphQL tools~(cf.\ Section~\ref{ssec:Background:ExistingTestSuites})
and some basic experimental results~\cite{DBLP:conf/tsp/RokselaKZ20},
	we observe that
there does not exist any methodological approach to thoroughly evaluate and compare the performance of approaches to create a GraphQL server. In this
	article
we introduce
	the Link\"oping GraphQL Benchmark~(\LinGBM)
to fill this gap.

\smallskip\noindent\textbf{Contributions and organization of the
	article:}
Our main contribution in this
	article
is \LinGBM, that is, a benchmark to experimentally study and compare the performance achieved by various approaches to create a GraphQL server. The benchmark consists of\footnote{All the material related to \LinGBM\ is available online (including, e.g., files with the query templates, the source code of tools, and documentation). In the related parts of this
	article
we
	provide links
to the relevant Web pages.} i)~a data schema for creating benchmark datasets at
	different scales,
ii)~16~query templates that cover different \removable{per\-for\-mance-re\-lat\-ed} challenges of GraphQL, and that can be used to create heterogeneous query workloads~(i.e., mixtures of diverse types of queries) for stress testing of systems and also to create various homogeneous workloads with specific types of queries for mi\-cro\-bench\-mark\-ing,
iii)~performance metrics and execution rules, and iv)~the necessary tooling to conduct experiments with the benchmark~(e.g.,
	a dataset \removable{generator, a} query workload generator,
test drivers).
Before describing these elements of the benchmark in detail~(Section~\ref{sec:artifacts}), we provide
	the relevant background on GraphQL and
on approaches to create GraphQL servers~(Section~\ref{sec:background}), and we detail the design considerations for the benchmark, including the design methodology, requirements, and design artifacts~(Section~\ref{sec:Design}).

Given the benchmark, we make further contributions: First,
	in Section~\ref{sec:Properties} we show statistical properties of the benchmark that are important to know when defining\hidden{ concrete} experiments. In particular, we measure dataset sizes and numbers of \removable{individual} queries per template, and we
		analyze result size distributions.
%
Thereafter, in Section~\ref{sec:Application} we demonstrate several
	mi\-cro\-bench\-mark\-ing
use cases in which we apply the benchmark. In particular, we show that the benchmark can be used i)~to evaluate the effectiveness of optimization techniques for GraphQL servers\hidden{ and how well they address specific \removable{per\-for\-mance-re\-lat\-ed} challenges that are relevant for building efficient GraphQL servers}, ii)~to study approaches that focus on improving the read scalability of GraphQL servers,
	and
iii)~to measure and compare the performance of GraphQL servers that are generated automatically by tools that provide such a functionality%
	.
In this context, we also present experimental results that highlight the pros and cons of selected techniques and tools, and we outline further application scenarios for the benchmark.

\section{Background} \label{sec:background}
This section provides a brief overview of GraphQL, including the relevant background on approaches to create GraphQL servers.

\subsection{Overview of GraphQL}
 \label{ssec:Background:OverviewOfGraphQL}

Conceptually, GraphQL consists of two core components: i)~a declarative language for clients to express data retrieval and data modification requests---this language is usually called the GraphQL query language---and ii)~another language that is used to define so-called GraphQL schemas, where for each GraphQL server, such a schema specifies the types of objects for which the server has data.
	Each such object consists of key-value pairs that are called \emph{fields} and that can be requested individually when retrieving data from the server; what fields an object has depends on the type of the object.
In the following, we first provide an overview of GraphQL schemas and, thereafter, highlight the main features of the query language.

\begin{figure}[t]
	\footnotesize%
\centering
\begin{tabular}{ll}
\begin{minipage}{.5\linewidth}%
\begin{verbatim}
        type Starship {
          name: String
        }

        type Character {
          name: String!
          appearsIn: [Episode]
          starships: [Starship]
        }
\end{verbatim}%
\end{minipage}
  &
\begin{minipage}{.5\linewidth}%
\begin{verbatim}
type Query {
  hero(ep:Episode!): Character
  allCharacters: [Character]
}

enum Episode {
  NEWHOPE  EMPIRE  JEDI
}

\end{verbatim}%
\end{minipage}
\end{tabular}
\caption{Small GraphQL schema with three object types and an enumeration type.}
\label{fig:SchemaExample}
\end{figure}

\subsubsection{GraphQL Schemas and the GraphQL Type System}
Figure~\ref{fig:SchemaExample} presents a small GraphQL schema which we use as an example. Each such schema
	defines
\emph{object types}
	such as \texttt{\small Starship}, \texttt{\small Character}, and \texttt{\small Query} in the example. The definition of each object type consists of field declarations
that define the fields that can be requested for objects of the given type. For instance, for any
	\texttt{\small Character} in our example, clients can request the fields \texttt{\small name}, \removable{\texttt{\small appearsIn}, and~\texttt{\small starships}}.

Every \emph{field declaration}
	has to include
the name of the field and the type of the values returned when the field is requested. Such a value type may be one of several predefined scalar types such as \texttt{\small String}, \texttt{\small Int}, and \texttt{\small ID}---in which case we call the field a \emph{scalar-typed field}---or it may be any of the types defined by the schema~%
	(e.g., the type of the \texttt{\small hero} field in the example is \texttt{\small Character})
or it may be a list type~(see, e.g., the field \texttt{\small starships} in the example). Additionally, each field declaration may contain declarations of arguments that can be passed by the clients in their requests. The meaning of such arguments is not made explicit in the schema but needs to be documented separately. However, a typical use of arguments is to enable clients to specify which value(s) they want to retrieve for the requested field. For instance,
	the \texttt{\small hero} field in the example 
has an argument named \texttt{\small ep} with which clients can specify an episode for which they want to retrieve the hero character.

A special object type that every GraphQL schema has to contain is the \texttt{\small Query}  type. The fields of this type are the possible starting
points for data retrieval requests~(see below). Another special object type, which is optional and not used in our example, is the \texttt{\small Mutation} type; the field declarations of this type define data modification operations that clients may issue. 
For more details on the GraphQL schema language we refer to the GraphQL specification~\cite{GraphQLSpec} as well as to Hartig and Hidders' formal definition of this language~\cite{DBLP:conf/grades/HartigH19}.

\begin{figure}[t]
	\centering
	\includegraphics[width=0.7\linewidth]{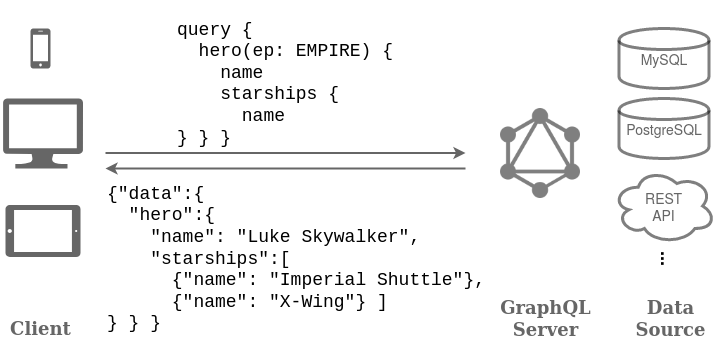}
	\caption{GraphQL querying scenario with an example query (above the arrows) and a
		response with a possible result for the query (below the arrows).}
	\label{fig:QueryExample}
\end{figure}

\subsubsection{Data Retrieval via the GraphQL Query Language}
Figure~\ref{fig:QueryExample} presents both
	an example query used for data retrieval based on
the schema in Figure~\ref{fig:SchemaExample} and a possible result for this query. 
As can be seen in this example,
	queries are
a form of nested expressions in which the most inner subexpressions are names of scalar-typed fields~(such as
	the two \texttt{\small name} fields in the example query);
for any other level of nesting, a field name is combined with an ordered set of subexpressions%
	\removable{~(e.g., \texttt{\small starships} in the example has one subexpression and \texttt{\small hero} has two)}%
.

The
	exact
way in which
	field names
can be used and nested within such a query is determined by the
	GraphQL schema of the GraphQL server.
At the
	root
level of the query, a field of the corresponding \texttt{\small Query} type must be used~(such as \texttt{\small hero} in the example query). If, according to the schema, the values of this field are objects~(i.e., it is not a scalar-typed field), then the fields of these objects may be used at the next level, and so on. Hence, every non-scalar-typed field of an object represents
	some kind
of pointer to other objects and can be used
	to request data about these objects within the same query (as demonstrated for \texttt{\small Starship} objects in the example query). 
%
	Moreover, fields in a query may be associated with arguments that are defined in the corresponding field declarations~(e.g., see the \texttt{\small hero} field).

The example also illustrates that the result of a GraphQL query
	follows the hierarchical structure of the query and contains the respective values for the requested fields and objects.
If the value type of a requested field is a list type, then the corresponding part of the query result is an array consisting either of scalar values or of objects with fields, depending on whether the list type is defined over a scalar type or an object type. Our example demonstrates the latter case based on the \texttt{\small starships} field. Also, while not shown in the example, objects in such arrays may again be nested.

There are a few more features in the GraphQL query language (namely, ``fragments'' and ``variables''). A complete definition of the language can be found in the GraphQL specification~\cite{GraphQLSpec}, and Hartig and P\'{e}rez provide a formal semantics for the language~\cite{Hartig18:GraphQLPaper}.

\subsection{Approaches to Create GraphQL Servers}
\label{ssec:Background:Approaches}

As an approach for building Web interfaces to access data in some underlying data sources, GraphQL is deliberately independent of what such data sources can be. Therefore, creating a GraphQL server for a particular application is usually a software engineering effort that involves
	the integration of relevant programming libraries and
the manual implementation of
	functionality to interact with the underlying data source(s)%
~\cite{Porcello:GraphQLbook, Grebe:GraphQLbook, Buna:GraphQLbook1, Buna:GraphQLbook2, Kimokoti:GraphQLbook, Williams:GraphQLbook}. On the other hand, for particular types of data sources~(such as specific database systems), there also exist tools that can be used to automatically create a GraphQL server on top of such a data source. In this section we provide
	a high-level
overview of these different approaches to build GraphQL servers, including some related optimization~techniques.

\subsubsection{Implementing Resolver Functions} \label{sssec:Background:Approaches:Resolvers}
When Facebook released the first public draft of the GraphQL specification, they also
	published a programming library\footnote{\url{https://github.com/graphql/graphql-js}} that provides a reference framework
for implementing GraphQL servers. The general implementation approach captured by this framework has become the prevalent way to build a GraphQL server, and it has also been adopted by several other frameworks for other programming languages. The main idea of this approach is to implement, for every field of every object type in the given GraphQL schema, a dedicated function---called \emph{resolver}---that can return the necessary data for this field for any object of the respective type. Then, for any given query that is valid for the implemented schema~(which is checked by the framework), the framework
	\removable{takes care of executing the query by using the resolvers. To this end, the framework} invokes the relevant resolvers recursively and assembles the query result based on their output.
By this approach, the developers of a GraphQL server have complete freedom in terms of how their resolvers fetch relevant data from whatever underlying data source(s) that they access. 

As a typical example of such resolvers, consider the GraphQL schema in Figure~\ref{fig:SchemaExample} and assume that the data for a GraphQL server with this schema is stored in an SQL database. The resolver for, say, the \texttt{\small hero} field may create an SQL query to fetch the ID, and perhaps also the name, of the
	character \removable{that is the} hero
of the episode given as argument. After obtaining these values (ID and name) from the database, the resolver returns an object consisting of this data.
	The resolvers for the fields of the type \texttt{\small Character} may expect such an object as input, which would be passed to them by the execution framework.
Then, the resolver for the \texttt{\small name} field \removable{of \texttt{\small Character}} can simply
	return the name value from such a given input object \removable{as its output}.
In contrast, the resolver for
	the \texttt{\small starships} field
may use the ID value to create an SQL query that fetches the names of the relevant starships and, then, return an array of objects with these names. For each object in this array, the execution framework can invoke the resolver for the \texttt{\small name} field of \texttt{\small Starship} if needed for the given~query.

While the example seems straightforward, things become less trivial
	when \removable{implementing}
GraphQL servers in which field arguments in the schema are used to capture filter or sorting conditions\footnote{\url{https://www.howtographql.com/graphql-js/8-filtering-pagination-and-sorting/}}
	for lists of objects or pagination\footnote{\url{https://graphql.org/learn/pagination/}} \removable{of such lists~(i.e., limit and offset)}.
Additionally, field values may not map directly to values
	available
explicitly in the underlying data source~(e.g., they may be aggregate values).
Different implementation strategies can be used and combined with one another to support all these features in a GraphQL server. Moreover, there exist different practices and tools to access possible data sources; for instance, instead of using SQL queries,  a GraphQL server may interact with an SQL database through an object-relational mapping.\footnote{\url{https://github.com/mickhansen/graphql-sequelize}}\footnote{\url{https://github.com/rse/graphql-tools-sequelize}} With the \LinGBM\ benchmark we aim to provide a well-designed testbed to evaluate these diverse practices, strategies, and tools in terms of their impact on the performance of a GraphQL server that uses them.

\subsubsection{Prominent Optimization Techniques} \label{sssec:Background:Approaches:Optimization}
In addition to the various general approaches to implement a GraphQL server based on resolver functions,
	several techniques
have been proposed to address specific performance pitfalls related to these approaches. Two prominent examples of such techniques are server-side batching and caching\footnote{\url{https://graphql.org/learn/best-practices/\#server-side-batching-caching}}\!, for which we show experimental results in this
	article
by~using~\LinGBM.

The idea of \emph{server-side caching} is to cache the response to every request that the resolvers make to the underlying data source, and if the exact same request is made again within the scope of executing a given GraphQL query, then use the response from the cache instead of accessing the data source again.
A typical approach to implement this technique is to use memoization.

The idea of \emph{batching} is to combine multiple similar requests to the underlying data source into a single request. Typically, this can be done for the requests issued by the same resolver based on different inputs. As an example, consider the following \removable{GraphQL}~query
	for the schema in Figure~\ref{fig:SchemaExample}.

\begin{center}%
\small%
\begin{verbatim}
       query{ allCharacters{ name starships{name} } }
\end{verbatim}%
\end{center}
%
During the execution of this query, the resolver for the \texttt{\small starships} field
is invoked once for every \texttt{\small Character} in the database. Each time, the resolver
	issues
the same kind of SQL query, just with a different character ID.
	These individual queries may be combined into a single SQL query, which avoids multiple round trips to the database server.
A popular tool to implement this form of request batching is called DataLoader\footnote{\url{https://github.com/graphql/dataloader}} which also supports server-side caching.


\subsubsection{Out-of-the-Box GraphQL Servers} \label{sssec:Background:Approaches:OutOfTheBox}
As an alternative to implementing the GraphQL server for an application manually, it is also possible to use a fully-automated solution. We distinguish two categories of such solutions: On the one hand, there are tools such as GRANDstack\footnote{\url{https://grandstack.io/}} that generate a GraphQL server together with an underlying database system by using a given GraphQL schema as input. Hence, such tools focus on applications for which a new database is needed that may be populated through the GraphQL API. In contrast, the second category
	consists of
\emph{generic GraphQL servers} such as
	Hasura\footnote{\url{https://hasura.io/}} and PostGraphile\footnote{\url{https://www.graphile.org/postgraphile/}}
that can be used out of the box to access an existing database via a GraphQL API. To this end, when starting up, these GraphQL servers examine the database schema and generate a corresponding GraphQL schema for it. This GraphQL schema is generated in such a way that every GraphQL query
	\removable{that is}
expressed in terms of this schema can be converted by the GraphQL server into a single query for the database~(e.g., a single SQL query). Hence, such generic GraphQL servers do not use the
	aforementioned re\-solvers-based execution approach that man\-u\-al\-ly-im\-ple\-ment\-ed GraphQL servers typically use.
%
	With \LinGBM\ we also aim to
provide a testbed for evaluating such generic GraphQL~servers.

\subsubsection{GraphQL Schema Delegation} \label{sssec:Background:Approaches:Delegation}
While generic GraphQL servers are convenient to use, a disadvantage of them is that the GraphQL schema they generate may not be in a form that is the most suitable one for the application clients. In such a case, it is possible to build another GraphQL server with a man\-u\-al\-ly-de\-fined, ap\-pli\-ca\-tion-spe\-cif\-ic schema such that this server
	\removable{uses the generic GraphQL server as its data source. Hence, this additional server}
rewrites incoming GraphQL queries\removable{~(which use the ap\-pli\-ca\-tion-spe\-cif\-ic schema)} into GraphQL queries
	\removable{to be forwarded} to
the generic GraphQL server. This architectural pattern of delegating the execution of queries, or parts thereof, from one server to another is called \emph{schema delegation} and there exist dedicated tools\footnote{e.g., \url{https://www.graphql-tools.com/docs/schema-delegation}} to implement such a delegating GraphQL server. Note that such schema delegation tools can be used not only for the use case outlined here but also, e.g., to build GraphQL gateways that integrate multiple GraphQL APIs.

\subsubsection{Other Approaches}
There certainly exist other ways to implement or create a GraphQL server. In this sense, the list of approaches outlined above is not meant to be exhaustive but to contain the most prevalent ones which also are relevant for our work in this~%
	article.

\begin{table}[t]
\caption{Comparison of existing GraphQL test suites.}%
\label{tab:testSuites}%
	\footnotesize%
\begin{tabular}{lcccc}
\hline
test suite / benchmark
& number \& size of datasets
& number of queries
& design method
\\ \hline
gbench\tnote{a} &
	1 (100 empty
		objects)* &
	5 queries& 
	unclear
\\
The Benchmarker framework\tnote{b} &
	1 (10 tuples)* &
	1 query& 
	unclear
\\
PostGraphile's GraphQL Bench\tnote{c} &
	1 (15,607 tuples)
	& 
	9 queries & 
	unclear
\\
Hasura's GraphQL Bench\tnote{d} &
	1 (23,288 tuples)
	&
	3 queries &
	unclear
\\
GraphQL server benchmark\tnote{e} &
	1 (60 tuples)
	& 
		4 templates, up to 10 instances each &
	unclear
\\
\LinGBM\ \emph{(our proposal)} &
		unbounded scale factor
	&
		16 templates, 100--1M instances each &
		choke-points
\\ \hline
\end{tabular}%
\begin{flushleft}
\footnotesize
$^\text{a}$\url{https://github.com/graphql-quiver/gbench}
\\
$^\text{b}$\url{https://github.com/the-benchmarker/graphql-benchmarks}
\\
$^\text{c}$\url{https://github.com/benjie/graphql-bench-prisma}
\\
$^\text{d}$\url{https://github.com/hasura/graphql-bench}
\\
$^\text{e}$\url{https://github.com/tsegismont/graphql-server-benchmark}
\\
*\emph{hardcoded in resolvers of the tested servers}
\end{flushleft}
\end{table}

\subsection{Existing Test Suites and Related Benchmarks}
\label{ssec:Background:ExistingTestSuites}

	While there is no work on performance benchmarks for GraphQL servers in the research literature, there
exist a few performance-related test suites~(cf.\ Table~\ref{tab:testSuites}). Essentially, these test suites are GraphQL variations of HTTP load testing tools such as
	wrk\footnote{\url{https://github.com/wg/wrk}} and vegeta\footnote{\url{https://github.com/tsenart/vegeta}}.
That is, each
	such test suite
consists of a specific dataset%
	~(of a comparably small size)%
, a few GraphQL queries, and a test
	driver that
records and visualizes
	latency or throughput
measurements obtained by issuing these queries to a GraphQL server built over~the~dataset.

We argue that these test suites are
	insufficient for benchmarking the performance achieved by different approaches to build GraphQL servers.
By focusing on a single
	(small)
dataset, these test suites cannot be used to study the behavior of GraphQL server implementations at scale. By using only a small number of fixed queries, it is not possible to extensively test or compare the throughput of systems that may apply caching on various levels. Additionally, it is not
	clear whether
the few selected queries test all important aspects of approaches to build GraphQL servers. 
Our work
addresses
	the limitations \removable{of the existing test suites}
and, more generally, the lack of a well-designed
	performance benchmark
for
	evaluating
and comparing approaches to build GraphQL~servers.

Given that
	GraphQL servers are not merely fetching content via canned queries from an underlying data source but, instead, employ some form of query processing themselves, 
our proposal
	combines the aspect of Web server benchmarking as captured by the existing GraphQL test suites with aspects of
benchmarks that test \removable{database} query engines.
	While there exists a plethora of \removableAltOff{such database}{such} benchmarks, the ones that are most related to our work are benchmarks that provide different tests where each test considers a specific type of queries or specific query features. A recent example is Lissandrini et al.'s mi\-cro-bench\-mark\-ing framework for graph databases which focuses on the performance of primitive operators~\cite{DBLP:journals/pvldb/LissandriniBV18}. Another example is WatDiv~\cite{DBLP:conf/semweb/AlucHOD14} which can be used to test the behavior of SPARQL query engines for queries that have specific structural features or specific data-driven features. Similarly, our benchmark consists of different templates for GraphQL query requests where each of them tests
		specific aspects of processing such requests~(as captured by the choke points in Section~\ref{ssec:Design:ChokePoints}).

Another form of benchmarks related to our work are
	\removable{so-called}
Web framework
	benchmarks such as the ones by TechEmpower\footnote{\url{https://www.techempower.com/benchmarks/}} and by The Benchmarker\footnote{\url{https://web-frameworks-benchmark.netlify.app/}}.
These benchmarks focus on the performance of specific tasks
	in back-end implementations of
Web applications when using different implementation frameworks\hidden{ for such applications}; such tasks include the serialization of data, responding to a single data retrieval request of a fixed type, responding to a sequence of such requests, update requests, etc. Our benchmark can be considered as a form of such Web framework benchmarks with a focus on frameworks for building GraphQL servers and, in particular, their their query processing capabilities.

\section{Design of the Benchmark} \label{sec:Design} 

The aim of our benchmark is to provide a framework that can be used to test and to compare the performance that can be achieved by different approaches to build GraphQL servers. To make this aim more concrete we identified two scenarios to be simulated by the benchmark~(cf.\ Section~\ref{ssec:Design:Scenarios}). Given these scenarios, we developed the benchmark by applying the
	design methodology for benchmark development
of the Linked Data Benchmark Council~\cite{DBLP:journals/sigmod/AnglesBLF0ENMKT14}. The main artifacts created by the process of applying this methodology are i)~a data schema, ii)~a workload of operations to be performed by the system under test, ~iii)~performance metrics, and iv)~benchmark execution rules. A crucial aspect of the methodology is to identify key technical challenges, so-called \emph{choke points}, for the types of systems for which the benchmark is designed. These choke points then inform the creation of the aforementioned artifacts.

In this section we first list key requirements
for a GraphQL performance benchmark; thereafter, we describe the two benchmark scenarios and provide an overview of the choke points to be defined for our benchmark. The actual benchmark artifacts shall then be introduced in
	Section~\ref{sec:artifacts}.

	\subsection{Requirements}
\label{sec:Requirements}

The following requirements \removable{that we deem important} for a GraphQL performance benchmark are derived from both the literature on similar benchmark projects~\cite{DBLP:journals/sigmod/AnglesBLF0ENMKT14, eichmann2020idebench, armstrong2013linkbench} and the aforementioned limitations of
	existing \removable{GraphQL}
test suites~(cf.\ Section~\ref{ssec:Background:ExistingTestSuites}).

\smallskip\noindent\textbf{R1:}
A GraphQL performance benchmark should contain workloads of operations~(queries and mutations) that capture all relevant features of GraphQL. The relevance of features depends on the use case \removable{scenario}(s) covered by a particular benchmark; without capturing all features relevant for
	the selected
scenario(s), a benchmark is incomplete.

\smallskip\noindent\textbf{R2:}
	A
performance benchmark should have a scalable dataset; that is, it should be possible to scale up the dataset to an arbitrary size by maintaining the main characteristics of the dataset (e.g., value distributions). This is important
	for experiments
to study
	how systems behave when facing increasingly larger datasets.

\smallskip\noindent\textbf{R3:}
Regarding metrics,
	a performance benchmark
should go beyond considering
	only throughput%
%
	. For instance, response times also directly impact user satisfaction.


\smallskip\noindent\textbf{R4:}
The amount of data
	to be processed
by each
	operation \removable{of the} workloads
should depend monotonically on the dataset size. For instance, for benchmark queries this means that their result size should not
	change back and forth
between greater and smaller numbers when executed over increasingly larger versions of the benchmark datasets. Otherwise, experiments may
	show
nondeterministic changes in the behavior of tested systems depending on which dataset scale~is~used.

\smallskip\noindent\textbf{R5:}
Like most traditional database benchmarks, a GraphQL benchmark must define default configurations for its datasets, workloads, and tests. However, as there is no one-size-fits-all configuration for all parameters of a benchmark, we argue that it is important to provide the ability to customize
	the parameters
to match different use cases.
	Of course, for publishing benchmark results, it then should be required to disclose the used configuration settings along with the benchmark results.

\smallskip\noindent\textbf{R6:}
The
	overall number of distinct operations~(e.g., queries) should be sufficiently high 
to
	support
extensive throughput experiments that require many disjoint sets of operations
	to be issued by different simulated clients.

\subsection{Scenarios} \label{ssec:Design:Scenarios}
Our benchmark focuses on
	the following
two
	types of use case scenarios.

\smallskip\noindent
\textbf{Scenario 1} represents cases in which data from a \emph{legacy database} has to be exposed as a \emph{read-only} GraphQL API with a \emph{user-specified GraphQL schema}.
Hence, this scenario focuses primarily on
	techniques and tools
to implement GraphQL servers manually~(cf.\ Sections~\ref{sssec:Background:Approaches:Resolvers} and~\ref{sssec:Background:Approaches:Optimization}). However,
	tools that can be used to create GraphQL servers but
that are not designed to support this scenario out of the box can also be tested in terms of this scenario. To this end, they have to be extended with an integration component such as a schema delegation layer~(cf.\ Section~\ref{sssec:Background:Approaches:Delegation}). In such a case, from the perspective of the benchmark, the combination of the integration component with the underlying
	GraphQL server created by the tool
are treated as a~black~box.

\smallskip\noindent
\textbf{Scenario 2} represents cases in which data from a \emph{legacy database} has to be exposed as a \emph{read-only} GraphQL API provided by an \emph{automatically generated GraphQL server}.
This scenario focuses on tools that auto-generate all artifacts necessary to set up a GraphQL API
	that provides access to
a legacy database, including the aforementioned generic GraphQL servers such as Hasura and PostGraphile~(cf.\ Section~\ref{sssec:Background:Approaches:OutOfTheBox}). Notice that such tools do not support the first scenario out of the box because any GraphQL API created by such a tool is based on a
	tool-spe\-cif\-ic generated
GraphQL schema~(%
	rather than
a us\-er-spec\-i\-fied one).

\begin{sidewaystable}
\caption{Choke points of the \LinGBM\ benchmark and their coverage by the 16 \LinGBM\ query templates.}%
\label{List-of-Choke-points}%
	\small
\centering
\begin{tabular}{ p{8mm}p{80mm}| cccc|cccc|cccc|cccc}
\hline
\multicolumn{2}{r|}{\textbf{QT:}} &
{\footnotesize 1}&%
{\footnotesize 2}&%
{\footnotesize 3}&%
{\footnotesize 4}&%
{\footnotesize 5}&%
{\footnotesize 6}&%
{\footnotesize 7} &
{\footnotesize 8} &
{\footnotesize 9} &
{\footnotesize 10} &
{\footnotesize 11} &
{\footnotesize 12} &
{\footnotesize 13} &
{\footnotesize 14} &
{\footnotesize 15} &
{\footnotesize 16} \\
\hline
\multicolumn{2}{l|}{\textbf{Attribute Retrieval}} &&&&&&&&&&&&&&&& \\
CP 1.1 & Multi-attribute retrieval & {\footnotesize$\times$} &  &  &  &  &  &  & {\footnotesize$\times$} &  & {\footnotesize$\times$} &  &  &  &  &  & {\footnotesize$\times$} \\ \hline
\multicolumn{2}{l|}{\textbf{Relationship Traversal}} &&&&&&&&&&&&&&&& \\
CP 2.1 & Traversal of \hidden{multiple }1:N relationship types                                 
& {\footnotesize$\times$} & {\footnotesize$\times$} &  & {\footnotesize$\times$} & {\footnotesize$\times$} & {\footnotesize$\times$} & {\footnotesize$\times$} &  & {\footnotesize$\times$} &  &  & {\footnotesize$\times$} & {\footnotesize$\times$} & {\footnotesize$\times$} &  & \\
CP 2.2 & Efficient traversal of 1:1 relationship types & {\footnotesize$\times$} &  & {\footnotesize$\times$} & {\footnotesize$\times$} & {\footnotesize$\times$} & {\footnotesize$\times$} & {\footnotesize$\times$} &  & {\footnotesize$\times$} &  & {\footnotesize$\times$} &  &  &  &  & \\
CP 2.3 &
	Relationship traversal
%
	with
retrieval of intermediate object data &  &  & {\footnotesize$\times$} & {\footnotesize$\times$} & {\footnotesize$\times$} &  &  &  &  &  & {\footnotesize$\times$} & {\footnotesize$\times$} & {\footnotesize$\times$} & {\footnotesize$\times$} &  &  \\
CP 2.4 & Traversal of relationships that form cycles &  &  &  &  & {\footnotesize$\times$} &  &  &  &  &  &  &  &  &  &  &  \\
CP 2.5 &
	Acyclic relationship traversal that visits \removable{data }objects repeatedly%
&  &  &  & {\footnotesize$\times$} &  & {\footnotesize$\times$} & {\footnotesize$\times$} &  & {\footnotesize$\times$} &  & {\footnotesize$\times$} &  & {\footnotesize$\times$} & {\footnotesize$\times$} &  &  \\ \hline
\multicolumn{2}{l|}{\textbf{Ordering and Paging}} &&&&&&&&&&&&&&&& \\
CP 3.1 & Paging without offset &  &  &  &  &  &  &  & {\footnotesize$\times$} & {\footnotesize$\times$} &  &  &  &  &  &  &  \\
CP 3.2 & Paging with offset &  &  &  &  &  &  & {\footnotesize$\times$} &  &  &  &  &  &  &  &  &  \\
CP 3.3 & Ordering  &  &  &  &  &  &  &  & {\footnotesize$\times$} & {\footnotesize$\times$} &  &  &  &  &  &  &  \\ \hline
\multicolumn{2}{l|}{\textbf{Searching and Filtering}} &&&&&&&&&&&&&&&& \\
CP 4.1 & String matching &  &  &  &  &  &  &  &  &  & {\footnotesize$\times$} &  &  & {\footnotesize$\times$} & {\footnotesize$\times$} &  & \\
CP 4.2 & Date matching &  &  &  &  &  &  &  &  &  &  &  &  &  & {\footnotesize$\times$} &  &  \\
CP 4.3 & Subquery-based filtering &  &  &  &  &  &  &  &  &  &  &  & {\footnotesize$\times$} & {\footnotesize$\times$} & {\footnotesize$\times$} &  &  \\
CP 4.4 & Subquery-based search &  &  &  &  &  &  &  &  &  &  & {\footnotesize$\times$} &  &  &  &  &  \\
CP 4.5 & Multiple filter conditions &  &  &  &  &  &  &  &  &  &  &  &  &  & {\footnotesize$\times$} &  &  \\ \hline
\multicolumn{2}{l|}{\textbf{Aggregation}} &&&&&&&&&&&&&&&& \\
CP 5.1 & Calculation-based aggregation &  &  &  &  &  &  &  &  &  &  &  &  &  &  &  & {\footnotesize$\times$} \\
CP 5.2 & Counting &  &  &  &  &  &  &  &  &  &  &  &  &  &  & {\footnotesize$\times$} &  \\ \hline
\end{tabular}
\end{sidewaystable}

\subsection{Choke Points} \label{ssec:Design:ChokePoints} 

As mentioned before, we have applied a choke-point based
	methodology%
~\cite{DBLP:journals/sigmod/AnglesBLF0ENMKT14} for designing our benchmark.
To this end, we have identified 16~choke points for GraphQL servers. As per our two benchmark scenarios~(which capture read-only use cases), these choke points focus only on queries. Table~\ref{List-of-Choke-points}~(left-hand side) lists these choke points, which are grouped into the following five classes.

\smallskip\noindent\textbf{Choke Points Related to Attribute Retrieval:}
Queries may request the retrieval of multiple
	scalar-typed fields
of the data objects selected by the queries. The
	\removable{technical}
challenge
	\removable{captured} by
the corresponding choke point is to
	fetch the values for these fields from the underlying data source using a single operation \removable{rather than performing a separate fetch operation for each~field}.
\hidden{%
An approach that is commonly used to
	address this challenge
is batching~(cf.\ Section~\ref{ssec:Background:Approaches}).%
}

\smallskip\noindent\textbf{Choke Points Related to Relationship Traversal:}
One of the main innovations of GraphQL in comparison to REST APIs is that it allows users to traverse the relationships between data objects in a single request. Supporting such a traversal in a GraphQL server may pose different challenges%
	, which are captured by the choke points in this class.
%
	For instance,
choke point~CP~2.4 captures the challenge to avoid unnecessary operations in cases in which relationships between requested objects form directed cycles.
	Queries that traverse along these relationships may come
back to an object that has been visited before on the same traversal path%
. A naive implementation may end up requesting the same data multiple times from the underlying data source. Even a more sophisticated solution that caches and reuses the results of such requests may end up repeating the same operations over the cached data.

\smallskip\noindent\textbf{Choke Points Related to Ordering and Paging:}
Given that an exhaustive traversal of a sequence of 1:N relationships may easily result in reaching a prohibitively large number of objects, providers of GraphQL APIs aim to protect their servers from queries that require such re\-source-in\-ten\-sive traversals. A common approach used in this context is to enforce clients to use paging when accessing 1:N relationships, which essentially establishes an upper bound on the maximum possible fan-out at every level of the traversal. A feature related to paging is to allow users to specify a particular order over the objects visited by traversing a 1:N relationship.
	This feature may be used in combination with paging, but also to simply request a particular order in which objects have to appear in the result.
This class of choke points focuses on implementing these features efficiently.

\smallskip\noindent\textbf{Choke Points Related to Searching and Filtering:}
Field arguments in GraphQL queries are powerful not only because they can be used as a flexible approach to expose paging and ordering features. Another use case, which is perhaps even more interesting from a data retrieval point of view, is to expose arbitrarily complex search and filtering functionality. The choke points in this class capture different challenges related to this use case.

\smallskip\noindent\textbf{Choke Points Related to Aggregation:}
Another advanced feature that GraphQL APIs may provide is to execute aggregation functions over the queried data. Challenges in this context are to compute aggregations efficiently~(CP~5.1)---e.g., by pushing their computation into the underlying data source---and to recognize that for counting, the corresponding objects/values may not actually have to be retrieved from the underlying data source~(CP~5.2).

\smallskip\noindent
For a detailed description of all 16~choke points covered by our benchmark we refer to the wiki%
	\footnote{\url{https://github.com/LiUGraphQL/LinGBM/wiki/Choke-Points}}
of the benchmark.

\section{Elements of the Benchmark} \label{sec:artifacts}

Now we are ready to introduce the different elements of \LinGBM.

\subsection{Data Schema} \label{ssec:Artifacts:DataSchema}

The data schema of the benchmark%
	\footnote{\url{https://github.com/LiUGraphQL/LinGBM/wiki/Data-Schema-of-the-Benchmark}}
consists of i)~a database schema for synthetic datasets that can be generated in the form of an SQL database or an RDF graph database, ii)~rules for generating such datasets in different sizes, iii)~a GraphQL schema for a GraphQL server that may provide access to any version of the benchmark dataset, and a iv)~a schema mapping that defines how the elements of the GraphQL schema map to the database schema.

\subsubsection{Datasets}
Instead of creating a new dataset generator from scratch, \LinGBM\ reuses the dataset generator of the Lehigh University Benchmark~(LUBM)~\cite{DBLP:journals/ws/GuoPH05}. LUBM is a popular benchmark in the Semantic Web community for evaluating the performance of storage and reasoning systems for RDF-based graph~datasets. 
%

\begin{sidewaysfigure}
	\centering
	\includegraphics[width=0.95\linewidth]{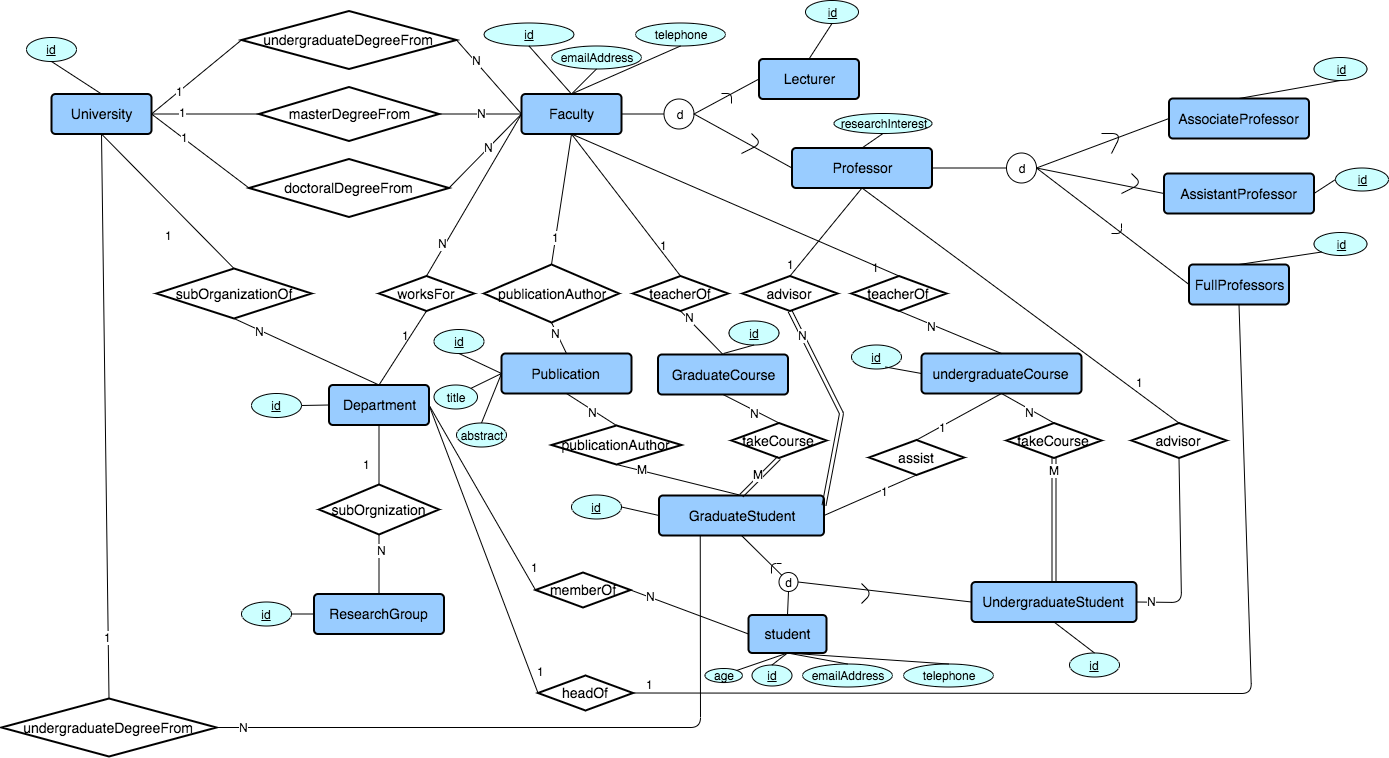}
	\caption{Conceptual schema of the LUBM datasets used for LinGBM (illustrated as an Entity-Relationship diagram).}
	\label{fig:LUBMschema}
\end{sidewaysfigure}

	Figure~\ref{fig:LUBMschema} illustrates the conceptual schema of these benchmark datasets which
capture a fictitious scenario of universities with departments, different types of faculty~(lecturers, assistant professors, etc.), students, courses, research publications, and
	\removable{other related types of entities as well as}
corresponding relationships between such entities. It is easy to imagine \removable{different} Web or mobile applications in such a scenario that enable students or researchers to browse and interact with the data, where these applications access the data via a GraphQL API. Hence, these datasets are a suitable starting point for a GraphQL~benchmark.

	In order ``\textit{to make the data[sets] as realistic as possible,}'' the dataset generator applies ``\textit{%
restrictions [that] are [...] based on common sense and domain investigation}''~\cite{DBLP:journals/ws/GuoPH05}. For instance,
	\removable{each} university has
15--25 departments,
	\removable{each department has 7--10 full professors,}
and the undergraduate student/faculty ratio per department is between 8 and~14, whereas the graduate student/faculty ratio
is between 3 and~4.\footnote{\url{http://swat.cse.lehigh.edu/projects/lubm/profile.htm}} The actual cardinalities
	\removable{of relationships}
are selected from these ranges uniformly at random. Similarly, when generating relationships between
	the entities created in an earlier stage of the data generation process,
the entities to be connected are selected uniformly at random from the corresponding pool of possible entities. Depending on the type of relationship, this pool of possible entities is either context specific~(e.g., students may take courses only from their department) or global~(e.g., grad students may have their undergraduate degree from any university).
The advantage of using uniform distributions for the data is that different queries of the same query template have the same predictable performance footprint; that is, they are roughly the same in terms of properties such as intermediate result sizes and overall result sizes~(as shall be confirmed in our analysis in Section~\ref{sec:sizeOfResult}; see, in particular, Figure~\ref{Fig:numberOfLeafnode}). 

In addition to being sufficiently realistic and diverse in terms of different types of relationships,
	\removable{another important property for our purposes is that}
these
datasets can be generated at different sizes where the number of universities to be created serves as the \emph{scale factor}. That is, the smallest dataset, at scale factor~1, consist of the data about one university.
	Yet another important property is that the data generation process is both deterministic and monotonic; hence,
all data that is generated at a smaller scale factor is guaranteed to be contained in every dataset generated with the same random seed at a greater scale factor.
Due to these properties,
	\removable{we consider the LUBM datasets as}
a suitable basis for our benchmark.
The fact that LUBM has been designed for a different purpose is not an
	issue in this context
because its focus on reasoning systems is reflected mainly in the queries defined for LUBM, not in~its~datasets.

	The only relevant limitation of the LUBM datasets is that they can be created only as RDF data. For
\LinGBM\ we wanted to also support SQL databases as underlying data sources for the tested GraphQL servers. Therefore, we have defined a relational database schema%
	\footnote{\url{https://github.com/LiUGraphQL/LinGBM/wiki/Datasets}}
that resembles the concepts and relationships of the LUBM ontology, and we have
	extended the dataset generator by implementing
a mapping from the
	generated RDF graphs
to
	\removable{SQL databases that are}
instances of our database~schema.

\subsubsection{GraphQL Schema and Schema Mapping}
In addition to the benchmark datasets, \LinGBM\ introduces a GraphQL schema for exposing any version of
	these datasets
as a GraphQL API. Essentially, this schema contains an object type for each type of entities in the benchmark dataset~(universities, departments, graduate students, etc). The fields of each such object type match both the attributes of the corresponding entity type and its relationships to other entity types. For example, the object type \texttt{\small GraduateStudent} in the \LinGBM\ GraphQL schema has fields such as \texttt{\small emailAddress} and \texttt{\small memberOf} where the former is for the email-address attribute of each graduate student in the generated datasets and the latter is for the
	membership
relationship that each such student has to
	the department they belong to%
. Hence, the value type of this \texttt{\small memberOf} field is the object type \texttt{\small Department} which, in turn, contains a field called \texttt{\small graduateStudents},
	\removable{with a \texttt{\small GraduateStudent} list as value type,}
to allow for GraphQL queries that traverse the relationship in the reverse direction.

In addition to
	the object types that we created by
this straightforward translation of the
	\removable{relational}
database schema into a GraphQL schema, we
	added a few more fields and types to the GraphQL schema
to be able to define queries that cover all of the aforementioned choke points of the benchmark. For example, some fields were extended with arguments to express filter conditions or requirements for sorting and paging.

	The complete \LinGBM\ GraphQL schema can be found online%
%
	\footnote{\url{https://github.com/LiUGraphQL/LinGBM/tree/master/artifacts}}%
\!, and we also
	provide a definition of
the
	exact
mapping%
	\footnote{\url{https://github.com/LiUGraphQL/LinGBM/wiki/Schema-Mapping}}
between this GraphQL schema and the schema of the benchmark datasets.
	We emphasize
that
	the \LinGBM\
GraphQL schema is relevant only for
	Scenario~1 of the benchmark%
	~(cf.\ Section~\ref{ssec:Design:Scenarios})%
.
GraphQL schemas as used in
	Scenario~2
are auto-generated by the corresponding systems~under~test.

\subsection{Query Templates}
\label{ssec:queryTemplate}

As a basis for creating query workloads, we have hand-crafted a mix of 16~templates of GraphQL queries
	such that\removable{, on one hand,} these queries
cover
	all the GraphQL-specific
choke points identified in the initial design phase of our benchmark~(cf.\ Section~\ref{ssec:Design:ChokePoints}). At the same time, given the university scenario represented by the benchmark datasets, the queries capture
	data retrieval requests that may be issued by Web or mobile applications built for
such a scenario.
We emphasize that these
	queries
are completely independent of the queries considered by the aforementioned LUBM benchmark. Although we adopt~(and extend) the dataset generator of LUBM, the queries of that benchmark are irrelevant for our purpose because they have been created with a focus on testing RDF-based storage and reasoning systems. In contrast,
	the mix of query templates that we have created
for \LinGBM\ focuses on GraphQL servers and their specific choke points.
Table~\ref{List-of-Choke-points} illustrates the coverage of these choke points by
	the~%
query~templates.

\begin{figure}[t]
\footnotesize
\begin{verbatim}
                              query qt5($departmentID:ID) {
                                 department(nr:$departmentID) {
                                    id 
                                    subOrganizationOf { 
                                       id
                                       undergraduateDegreeObtainedBystudent {
                                          id 
                                          emailAddress 
                                          memberOf { 
                                             id 
                                             subOrganizationOf {
                                                id 
                                                undergraduateDegreeObtainedBystudent {
                                                   id 
                                                   emailAddress 
                                                   memberOf { id }
                               } } } } } } }
\end{verbatim}
\caption{Query template QT5.}
\label{fig:QT5}
\end{figure}

Each such template is a GraphQL query that contains \removable{at least one} placeholder for
	specific
values that exist in the generated benchmark datasets. Hence, to instantiate
	any of the templates
into an actual query, every placeholder has to be substituted by one of the possible values. While all 16~templates can be found online%
	\footnote{\url{https://github.com/LiUGraphQL/LinGBM/tree/master/artifacts/queryTemplates}}%
\!,
	including a detailed description of each of them%
		\footnote{\url{https://github.com/LiUGraphQL/LinGBM/wiki/Query-Templates-of-the-Benchmark}}\!,
in the following, we describe
	two of them as exemplars.

Figure~\ref{fig:QT5} presents query template QT5, which is a typical example of queries that traverse relationships in cycles and that, thus, may come back to the same objects multiple times. In the particular case of QT5, the traversal starts from a given department, retrieves the university of this department, then proceeds to retrieve all graduate students with an undergraduate degree from this university, and then to the departments that these students are members of.
	This cycle is repeated
two times. 
Hence,
	this query template
covers choke point~CP~2.4. Additionally, by requesting
	the students' email addresses along the way,
the template also covers choke point~CP~2.3. Furthermore, the template covers CP~2.1~(because of the traversal from a university to graduate students) and CP~2.2~(because of the traversal from departments to their respective university, as well as from each graduate student to their department). The placeholder of this query template is \texttt{\small \$departmentID}, which is used to select a department based on its number (i.e., the \texttt{\small nr} attribute) as a starting point for the traversal. Hence, for any benchmark dataset,
	the number of every department
in this dataset can be used
	to instantiate
QT5 in order to obtain queries that can be used for the dataset, as well as for all datasets generated with scale factors greater than the given~dataset.%
\todo{TODO: If time (and space) permits, emphasize that these templates are only for Scenario~1, and they have to be translated for the auto-generated schema(s) in Scenario~2. There is some text for this already in Section~\ref{subsec:UC3}.}

	Queries of the second example template, QT9 in Figure~\ref{fig:QT9},
retrieve data about the publications of the advisors of 50 graduate students in a given university. For each such advisor, the publications have to be sorted on a given field. 
This query template contains two placeholders, \texttt{\small \$universityID} and \texttt{\small \$attrPublicationField}, where the former expects the ID of some university, and the latter expects a field name of publication objects that will be used for determining the sort order. Any pair of a possible university ID and a field of publication objects can be used to instantiate this template. The choke points that are covered by this template are CP~3.1~(because the template uses paging without offset for the graduate students) and CP~3.3~(because it requires sorting of the publications). Additionally, CP~2.1 is covered (by the traversal from a university to its graduate students, and from advisors to their publications), and so is CP~2.2 (by the traversal from the graduate students to their respective~advisor).


\begin{figure}[t]
\footnotesize
\begin{verbatim}
              query qt9($universityID:ID, $attrPublicationField:PublicationField) { 
                 university(nr:$universityID) { 
                    undergraduateDegreeObtainedBystudent(limit:50) { 
                       advisor { 
                          publications(order: {field:$attrPublicationField, direction:DESC}) { 
                             id 
              } } } } }
\end{verbatim}
\caption{Query template QT9.}
\label{fig:QT9}
\end{figure}

\subsection{Performance Metrics} \label{sec:Metrics}

The performance metrics considered by \LinGBM\ are defined based on the following three notions:
\begin{itemize}
	\item \emph{Query execution time} (\emph{QET}) is the amount of time~\removable{(typically given in milliseconds\removable{, but may also be specified otherwise})} that passes from the begin of sending a given query to the GraphQL server under test until the complete query result has been received in~return.

	\item \emph{Query response time} (\emph{QRT}) is the amount of time~\removable{(typically given in milliseconds)} that passes from the begin of sending a given query to the GraphQL server under test until the begin of receiving the query result in~return.

	\item \emph{Throughput} is the number of queries that are processed completely by a GraphQL-based client-serv\-er system within a specified time interval, where a query is considered to be processed completely after its complete result has been received by the client that requested the execution of the~query.
	\removable{By default, the time intervals considered have a duration of 60~seconds; however, the interval duration may also be specified differently for a specific~experiment.}
\end{itemize}

\noindent
Then, for \textbf{single queries}, we define the following metrics:

\begin{itemize}
	\item \aQETq\ is the average of the QETs measured when executing an individual query multiple times with the GraphQL server under test. When reporting this metric, the corresponding standard deviation has to be reported as well.
	\item \aQRTq\ is the QRT-specific counterpart of \aQETq.
\end{itemize}

\noindent
For whole \textbf{query templates}, we define the following metrics:

\begin{itemize}
	\item \QETt\ is the distribution of the individual QETs measured for multiple queries of the same template.
	\item \aTPt\ is the average of the throughput measured when running the same query workload multiple times, where the
		queries in the workload are all from the same
	template.
\end{itemize}

\noindent
For \textbf{mixed workloads}
	with queries from multiple templates, we define the following metrics:

\begin{itemize}
	\item \aTPw\ is the average of the throughput measured when running the same mixed query workload multiple times.
	\item \aTPm\ is the average of the throughput measured for multiple mixed workloads, where each such workload is run once.
\end{itemize}

\subsection{Tools} \label{ssec:Artifacts:Tools}
This section describes the tools that we have developed to enable users to perform experiments with the benchmark.

\subsubsection{Dataset Generator}
The \LinGBM\ dataset generator is an extension of the LUBM dataset generator. While the latter generates the benchmark datasets in the form of an RDF graph, our extension additionally supports generating SQL databases. The RDF versions can be written in several RDF serialization formats,
and the SQL-database versions are written as an SQL dump file that can be imported by a MySQL server or by a PostgreSQL server currently. In addition to
	the file with
the actual dataset, a corresponding metadata file is written. This extra file contains relevant values such as IDs of all generated entities and lists of words in generated publication titles and abstracts. These values are used by the query generator when instantiating query templates.

\subsubsection{Query Generator}
This tool generates actual queries by instantiating the query templates of the benchmark. To
	this end, the placeholders in a given
template are replaced by actual values. These values are selected uniformly at random from the metadata for a particular benchmark dataset. The resulting query instances can then also be used for any benchmark dataset generated at a greater scale factor%
	\removable{~(because the data generation process is monotonic as mentioned above)}%
.

When generating queries, users can specify the number of instances they want for each query template, and the query generator makes sure that the set of generated instances is duplicate-free. However, there are cases in which the dataset metadata at the given scale factor does not contain enough possible values to create the requested number of instances for some template%
	\removable{~(cf.\ Section~\ref{sec:nrOfInstance})}%
. In these cases, the
	query generator
creates as many instances as there are values.
In any case, every generated query instance is also assigned an identifier that is unique within the set of generated instances.

\subsubsection{Test Driver for Throughput Experiments}
This tool can be used to measure
	throughput-related metrics \removable{such as \aTPt, \aTPw, and \aTPm}.
%
To this end, the tool simulates
	one or more
clients that concurrently send queries
	to the GraphQL server under test. These queries can either be all from the same template~(in experiments that focus on \aTPt) or from a given mixed workload~(as considered for \aTPw\ or \aTPm).
%
Internally, the tool employs multi-threading with a special thread for coordinating the simulated clients. This coordination thread controls what is done and provides the
	\removable{simulated}
clients with sequences of test queries. 

When starting the test driver, the coordination thread first creates clients and distributes queries according to the specified number of clients.
%
Then, each client sends its queries sequentially%
%
	; that is, after receiving the result of a query, the next available query is sent. Additionally, before moving on to the next query, the client also
sends a measurement record to the coordination thread. Such a record consists of the identifier of the query, the QET, and an error code
	if the tested server responded with an error message.
Once a client has reached the end of its sequence of queries, it starts again from the beginning of the sequence. 

This process continues until a specified amount of time has passed~(60%
	~seconds
by default).
At this point, the coordination thread terminates all
	\removable{simulated}
clients and writes the collected measurement records into
	a CSV file and, then, moves on to restart the process for either the next template or the next given mixed workload (if any).
Additionally, the coordinator writes an extra file
	to record the total number of successful and unsuccessful queries per template/workload.
The overall process can be repeated to obtain multiple such measurements per template/workload
	to calculate
an average~(\aTPt,~\aTPw).

\subsubsection{Test Driver for QET and QRT}
This tool can be used to measure query execution times and query response times. For a specified set of query templates~(which may be all the templates of the benchmark) with a given set of query instances for each of them, the tool picks queries from the templates in a round-robin fashion and sequentially sends these queries to the server under test. That is, it first uses the first query instance of each template, one after another. After reaching the last template, it starts again from the first template and uses the second query instance, etc.
\todo{During the execution time and response time testing, in order to reduce the impact between queries, we set a waiting time between the queries. Specifically, when the master receives the response of a query, the next query is issued after waiting 1 second.}
This process continues until all given queries of all considered templates have been executed and, thus, QET and QRT have been recorded for each of the queries. The individual QET measurements can be combined into a \QETt\ distribution per template.

By doing multiple runs of such a measurement process, the tool can be used also to obtain the measurements for the single-query metrics \aQETq\ and \aQRTq. Currently, for such repeated runs, the
	test driver
tool has to be started repeatedly from within a separate~script.

\section{Properties of the Benchmark} \label{sec:Properties}
Defining concrete experiments based on a benchmark often requires an understanding of statistical properties of the
	benchmark.
Therefore, in this section, we show such properties for our benchmark. In particular,
	we report on the actual dataset sizes at the different scale factors, and
we analyze how the number of instances per query template scales with the scale factor and how the results of these query instances vary in terms of
	their size.

\subsection{Dataset Size at Different Scale Factors}\label{ssec:Properties:DatasetSize}

The size of each benchmark dataset depends on the corresponding scale factor. While this size may be measured using different metrics, we focus on
	three such metrics:
i)~the file size of the generated SQL import scripts, ii)~the sum of the number of rows across all tables of the generated SQL database, and iii)~the sum of the overall number of objects for all types of the \LinGBM\ GraphQL schema. Table~\ref{tab:sizeOfDataset} presents these statistics for the datasets
	generated
at scale factors 1, 10, 20, 100, and 150.
	Given these figures, we observe that the dataset size (in terms of each of the three metrics) increases linearly with the scale factor.

\begin{table}[b]
\centering
\caption{Sizes of the benchmark datasets at different scale factors.}%
\label{tab:sizeOfDataset}%
\small%
\begin{tabular}{lccccccc}
\hline
 & $\sfactor=1$ & $\sfactor=5$ & 
 $\sfactor=10$ & $\sfactor=15$ & 
 $\sfactor=20$ & $\sfactor=100$ & $\sfactor=150$ \\ \hline
file size & 12~MB & 78~MB & 
161~MB & 247~MB & 
340~MB & 1.66~GB & 2.60~GB \\
overall rows 
& 43,319 & 266,267 & 
542,467 & 832,142 & 
1,145,002 & 5,707,958 & 8,490,274 \\
overall objects 
& 17,195 & 102,368 & 
207,426 & 318,319 & 
\phantom{1,}437,555 & 2,179,766 & 3,243,523 \\
\hline
\end{tabular}%
\end{table}


	\subsection{Number of Query Instances at Different Scale Factors} \label{sec:nrOfInstance}

	As mentioned before, each of the \LinGBM\ query templates can be instantiated into actual queries by substituting the placeholder(s) of the template with one of the corresponding values from the benchmark dataset. Therefore, for
each query template, the number of possible instances of the template depends on the number of possible values for its placeholder(s), and this number, in turn, may depend on the scale factor~(bigger versions of the benchmark datasets may contain more possible values). Table~\ref{tab:Scale-NumberOfInstance-QT} lists these numbers
	\removable{for each template}
at different scale~factors.

\begin{table}[t]
\centering
	\small
\caption{Number of query instances of each query template at different scale factors.}
\label{tab:Scale-NumberOfInstance-QT}
\begin{tabular}{@{}c|rrrrr}
\hline
& $\sfactor=1$ & $\sfactor=5$ & $\sfactor=10$ & $\sfactor=15$ & $\sfactor=20$ \\ \hline
QT1 & 540 & 3,373 & 6,843 & 10,521 & 14,457 \\
QT2 & 1,000 & 1,000 & 1,000 & 1,000 & 1,000 \\
QT3 & 224 & 1,371 & 2,827 & 4,407 & 6,032 \\
QT4 & 93 & 562 & 1,128 & 1,745 & 2,399 \\
\hline
QT5 & 15 & 93 & 189 & 293 & 402 \\
QT6 & 1,000 & 1,000 & 1,000 & 1,000 & 1,000 \\
QT7 & 48,950 & 246,500 & 493,250 & 740,400 & 989,250 \\
QT8 & 15,000 & 15,000 & 15,000 & 15,000 & 15,000 \\
\hline
QT9 & 2,000 & 2,000 & 2,000 & 2,000 & 2,000 \\
QT10 & 27,077 & 170,576 & 344,750 & 530,133 & 728,208 \\
QT11 & 1,000 & 1,000 & 1,000 & 1,000 & 1,000 \\
QT12 & 14,685 & 458,490 & 1,864,485 & 4,338,744 & 7,953,570 \\
\hline
QT13 & 899,701 & 27,893,940 & 113,921,020 & 262,560,648 & 480,241,305 \\
QT14 & 6,297,907 & 195,257,580 & 797,447,140 & 1,837,924,536 & 3,361,689,135 \\
QT15 & 1,000 & 1,000 & 1,000 & 1,000 & 1,000 \\
QT16 & 1,000 & 1,000 & 1,000 & 1,000 & 1,000
\\ \hline
\end{tabular}
\end{table}

	For
most templates, the number of instances increases with the scale factor. For instance, by the discussion of QT5 in Section~\ref{ssec:queryTemplate}, we know that QT5 can be instantiated based on every department ID in the generated datasets. Since each university in the benchmark datasets has 15--25 departments, the number of instances of QT5 at scale factor~$\sfactor$ is, thus, between $15\!\cdot\!\sfactor$~and~$25\!\cdot\!\sfactor\!$.

For templates QT2, QT6, QT8, QT9, QT11, QT15, and QT16, the number of instances is independent of the scale factor. 
For instance, QT9 has 2,000~instances at every scale factor%
	\removable{, which can be explained as follows}.
Recall from Section~\ref{ssec:queryTemplate} that QT9 contains the placeholders \texttt{\small \$universityID} and \texttt{\small \$attrPublicationField}, where the number of possible values for the former is 1,000 at any scale factor%
	\hidden{~(including any $\sfactor > \text{1,000}$!)}
, and for the latter, only two fields of publication objects are considered as possible values~(namely, \texttt{\small title} and \texttt{\small abstract}).

\removable{
Query template~QT8 is another special case.
	It
has three placeholders, \texttt{\small \$cnt}, \texttt{\small \$attrGStudent1}, and \texttt{\small \$attrGStudent2}, where \texttt{\small \$cnt} specifies a limit for paging of a list of objects and the value of this parameter is selected randomly from the interval (500, 1000]. The values of the other two placeholders are the names of two distinct
	scalar-types fields
of the \texttt{\small Graduate\-Student} type and they specify a primary and a secondary sort order for a requested list. Since there are six such
	fields,
the number of instances of QT8 is $500\!\cdot\!6\!\cdot\!5=\text{15,000}$.%
}


	\subsection{Query Result Sizes at Different Scale Factors}\label{sec:sizeOfResult}


\begin{figure}[b]
  \centering%
  \subfloat[QT5]{\label{figur:QT5_leafNode}\includegraphics[width=0.4\linewidth]{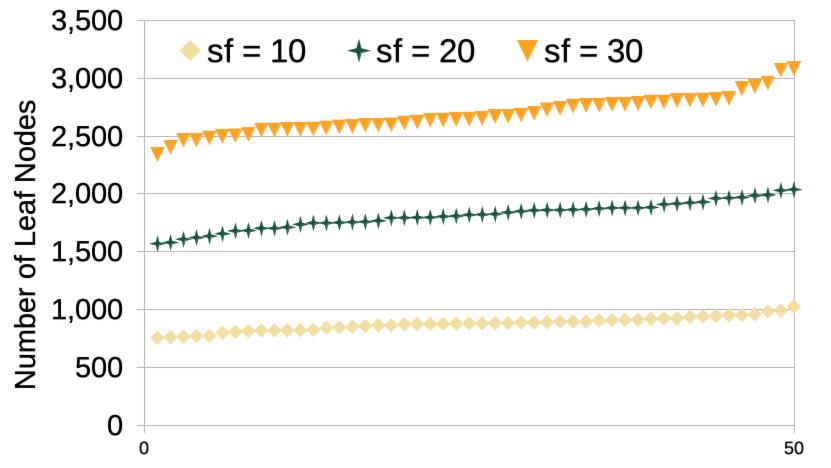}}%
\quad
  \subfloat[QT9]{\label{figur:QT9_leafNode}\includegraphics[width=0.4\linewidth]{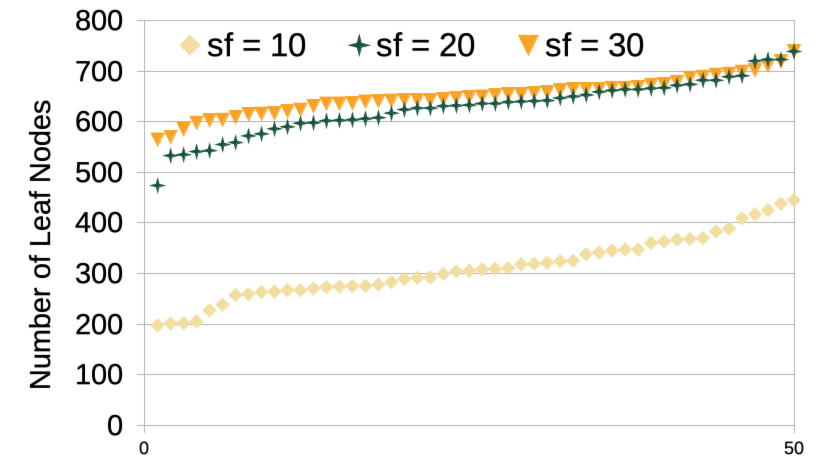}}%
  \caption{Result size distributions at different scale factors for 50 randomly selected queries of templates
		QT5 and QT9.
	The queries are sorted on the x-axes
		in increasing order of
	their result~sizes.}
  \label{Fig:numberOfLeafnode}
\end{figure}

Given that the execution time of a query may depend on the result size, result size distributions are an important statistic of a performance benchmark.
	To achieve an understanding of such distributions in our benchmark
we randomly selected 50~queries of each template and executed them over multiple, increasingly bigger datasets.
	\removable{As an example,}
Figure~\ref{Fig:numberOfLeafnode} illustrates the
	obtained result size distributions for query templates~QT5 and QT9, where we measure the size of each query result in terms of the number of leaf nodes in its tree representation.

Additionally, Table~\ref{tab:estimated-no-of-leaf-nodes} lists the minimum and the maximum values for these sizes for all possible instances of every
	\removable{query}
template. These lower and upper bounds can be calculated based on the value distributions used by the dataset generator.
\todo{TODO: either mention that the table also lists the average results sizes at scale factors 10 and 20 or remove these things from the table again; note that the first option requires discussing these numbers a bit!}
For instance, the queries of template~QT5 traverse along an N:1 relationship~(from the selected department to the university of the department), followed by a 1:N relationship~(from the university to the graduate students who obtained their undergrad degree from that university) and another 1:N relationship~(to the departments that these graduate students belong to); thereafter, for every department object reached by this traversal, the query traverses again along these three types of relationships. The value of~N in the second step can range from 0 to $7\!\cdot\!\sfactor$~(for every university in the generated datasets, there are at most $7\!\cdot\!\sfactor$ grad students with their undergrad degree from that university). Consequently, the number of leaf nodes in the query results can range from 0 to $(7\!\cdot\!\sfactor)^2$.

%

\begin{table}[t]
\centering
\small
	\caption{Bounds for number of leaf nodes in query results, depending on scale factor.}%
\label{tab:estimated-no-of-leaf-nodes}%

\begin{tabular}{@{}c|rr}
\hline
& min & max \\ \hline
QT1 & 0 & 7$\cdot\sfactor$  \\
QT2 & 0 & 80$\cdot\sfactor$ \\
QT3 & 1 & 1 \\
QT4 & 0 & 7$\cdot\sfactor$ \\
\hline
\end{tabular}
\quad
\begin{tabular}{@{}c|rr}
\hline
& min & max \\ \hline
QT5 & 0 & 49$\cdot\sfactor^2$ \\
QT6 & 0 & 7$\cdot\sfactor$ \\
QT7 & 0 & 10 \\
QT8 & 500 & 1000 \\
\hline
\end{tabular}
\quad
\begin{tabular}{@{}c|rr}
\hline
& min & max \\ \hline
QT9 & 0 & 1000 \\
QT10 & 0 & 6300$\cdot\sfactor$ \\
QT11 & 0 & 7$\cdot\sfactor$ \\
QT12 & 0 & 840 \\
\hline
\end{tabular}
\quad
\begin{tabular}{@{}c|rr}
\hline
& min & max \\ \hline
QT13 & 0 & 21$\cdot\sfactor$ \\
QT14 & 0 & 21$\cdot\sfactor$ \\
QT15 & 1 & 1 \\
QT16 & 1 & 1 \\
\hline
\end{tabular}
\end{table}


Given the ranges in Table~\ref{tab:estimated-no-of-leaf-nodes}%
	\ and the corresponding result size distributions~(e.g., Figures~\ref{figur:QT5_leafNode} and~\ref{figur:QT9_leafNode})%
, we observe that the 16~query templates can be classified into three groups: One group
	consists of
the ten templates for which the query result sizes depend on the scale factor~(QT1, QT2, QT4--QT6, QT10, QT11, QT13, QT14). Given the corresponding result size distributions~(e.g., Figures~\ref{figur:QT5_leafNode} and~\ref{figur:QT9_leafNode}), we observe that i)~for each query of these templates, the result sizes actually increase monotonically with an increasing scale factor and ii)~at each scale, different queries of the same template have results of different sizes.
The latter also holds for query templates~QT7--QT9 and~QT12---which form a second group---but not for the remaining templates that then belong to the third group~(QT3, QT15, and QT16). In fact, all queries of any such third-group template have a result size of~1~(at all scales). The reason in the case of QT3 is that QT3 queries traverse only along N:1 relationships, whereas for QT15 and QT16, the query results simply are single objects with aggregation values. Like for these third-group templates, the queries of the second-group templates\removable{~QT7--QT9 and~QT12} have results that are independent of the scale factor; in this case, the reason is that these queries use paging\removable{~(QT7--QT9)} or a specific filter\removable{~(QT12)} such that the result sizes are limited to a fixed upper bound. In the case of paging, this upper bound may become relevant only above a specific scale factor as can be seen for QT9 in Figure~\ref{figur:QT9_leafNode}.

This analysis shows that, in addition to covering all choke points, our mixture of query templates is also diverse in terms of result size characteristics%
	. Hence, the query templates
can be used to form heterogeneous workloads for stress testing of systems; and the templates can also be used to form various homogeneous workloads to test specific choke points as well as specific scaling~behavior.

\section{Application of the Benchmark} \label{sec:Application}

In this section we demonstrate the applicability of \LinGBM\ for
	three
different
	mi\-cro\-bench\-mark\-ing
use cases and present corresponding experimental results.
We begin with a description of the experimental setup that we have used as a basis for all the experiments.


\subsection{General Experiments Setup} \label{sec:configuration}

All experiments described in the following have been performed on a server machine with two 8-core Intel Xeon E5-2667 v3@3.20GHz CPUs and 256~GB of RAM. The machine runs a 64-bit Debian GNU/Linux~10 server operation system.
On this machine, we use Docker~(v9.03.6) to run all components of the experiment setups in a separate, virtual environment~(e.g., the GraphQL server under test, the database server used as data source, and the \LinGBM~test~driver).

All
	variations of
GraphQL servers that we have implemented
	manually
for the experiments are node.js~(v10.21.0) applications that use the Apollo Server package~(v2.17.0) and, for database access, the knex.js package~(v0.20.15).
%
As database server we use PostgreSQL~(v12.1, default configuration options only), given as a public Docker image, for which we limit the available resources to two vCPUs
and 1~GB of virtual~RAM.
%
%
To obtain the relevant measurements we used the corresponding \LinGBM\ test drivers~(cf.\ Section~\ref{ssec:Artifacts:Tools}).


Based on some preliminary tests with this setup, we selected the following default parameters for the experiments. Unless specified otherwise, we use scale factor~100 and, to connect to the database server, the man\-u\-al\-ly-im\-ple\-ment\-ed GraphQL servers use connection pooling with
up to 10~%
	\removable{parallel}
connections.
%
	Most of
our experiments focus on
	average throughput per template~(\aTPt,
cf.\ Section~\ref{sec:Metrics}) \removable{with one client},
	for which
we always do six runs of 60~seconds where the first run is regarded as a warm-up and the number of successfully completed queries per each of the other five runs are averaged. To have a sufficiently high number of distinct queries for these throughput runs, we generated 5000 queries for every query template for which this is possible at scale factor~10~(which is the smallest scale factor we used in one of our experiments), and for the other templates we used the maximum possible~(cf.\ Table~\ref{tab:Scale-NumberOfInstance-QT}).
%
Our preliminary tests also showed that
	there is only a marginal difference between query execution times and query response times~(\aQETq\ vs.\ \aQRTq).
This is because all tested
	GraphQL
servers return results only after having produced them completely; producing and returning results in a streaming manner is an open problem for GraphQL servers. Consequently, we ignore \aQRTq\ in the experiments.

\subsection{Evaluation of Optimization Techniques} \label{subsec:UC1}
The aim of our first use case is to evaluate the effectiveness of the two \removable{aforementioned} optimization techniques---serv\-er-side batching and caching%
	\removable{~(cf.\ Section~\ref{sssec:Background:Approaches:Optimization})}%
---and to
	achieve \removable{an understanding of} the choke points that
%
	\removable{each of} them
%
	\removable{can help to}
address. To this end, we have developed a GraphQL server for Scenario~1 of the benchmark
	\removable{(cf.\ Section~\ref{ssec:Design:Scenarios})}
by using a straightforward, re\-solver-based implementation. This server represents the baseline for the evaluation and we call it the \emph{naive} server. Thereafter, we have extended this server in three different ways to obtain three additional
	variations of
test servers: As a first variation, we have integrated serv\-er-side caching using memoization. For the second variation, we have replaced the naive resolvers by resolvers that use DataLoader to implement both serv\-er-side caching and batching. The third variation is a version of the second with caching disabled in DataLoader~(i.e., it uses only batching). The source code for these four test servers~(the naive one plus the three extended variations) is available online.%
	\footnote{\url{https://github.com/LiUGraphQL/LinGBM-OptimizationTechniquesExperiments}}

\subsubsection{Initial Macro-Level Comparison} \label{sssec:UC1:MacroLevel}
We begin by comparing the test servers based on the \aTPm\ metric using six different mixed workloads. Each of these workloads is a
	different, randomly sorted sequence of the same
100 queries from each template~(i.e., 1600 queries per workload in total). We have measured \aTPm\ with one client and a runtime of 600~seconds per workload.\footnote{We use a longer duration for these runs
	(600~seconds rather than 60~seconds)
to ensure that the tested servers have to process a greater selection of queries from each template.} For each tested server, the first workload was used as a warm-up and the throughputs achieved for the other five workloads were averaged to calculate \aTPm, which gives us the following measurements.

The naive server achieved an
	\aTPm\
of 200~queries~(with a standard deviation of~$\pm 38.84$); for the server with caching, it is 312~($\pm 15.41$); with batching, 729~($\pm 33.39$); and with both batching and caching together, 735~($\pm 32.77$). These numbers show that i)~both optimization techniques improve upon the baseline of the naive server, ii)~batching is significantly more effective than caching, and iii)~adding caching to batching does not lead to a significant improvement over batching alone.

While this experiment, with its diverse mix of queries, gives us a general idea of the
	improvements that may be obtained by using
the two optimization techniques, it does not allow us to derive more detailed insights about them. To gain such insights we can leverage the tem\-plate-lev\-el and query-lev\-el metrics of \LinGBM\ as
	demonstrated in the following.

\subsubsection{Experiments for a Micro-Level Comparison}
\begin{figure}[t]
\centering
\includegraphics[width=\linewidth]{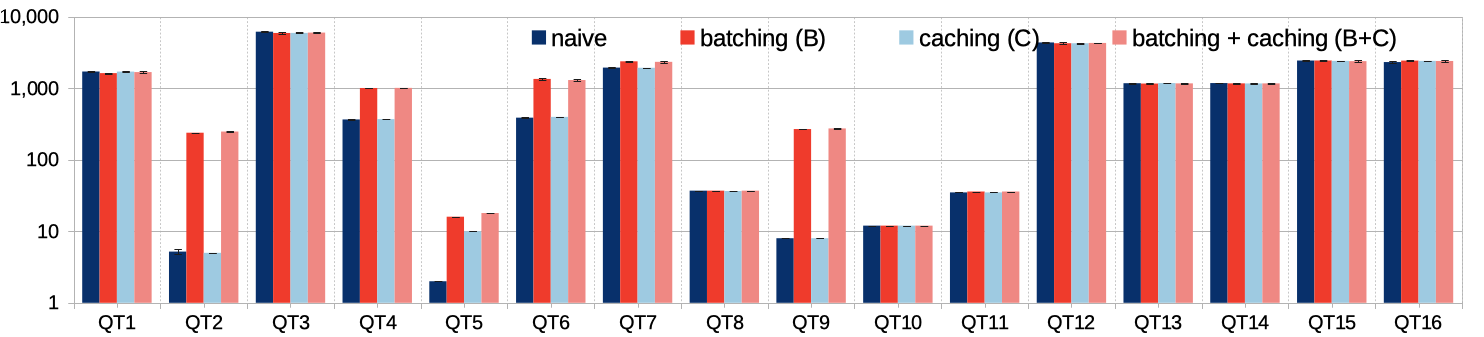}
\caption{Comparison of the manually-implemented GraphQL servers in terms of average throughput
	per template~(\aTPt)
with one client at scale factor 100.}
\label{Fig:TP_oneClient}
\end{figure}

As a first
	microbenchmarking
experiment, we have measured the
	\removable{average throughput per query template}~(\aTPt)
that
	each of the four test servers achieves
for
	\removable{the queries of}
each of the 16~\LinGBM\ query templates at scale factor~100 with one client. 
Figure~\ref{Fig:TP_oneClient} illustrates these measurements%
	~(error bars in the bar charts represent one standard deviation for the corresponding~averages)%
.
Thereafter, we have measured the corresponding \QETt\ required by the test servers for a single run with 100~randomly selected queries per template. The box plots in
	Figures~\ref{figur:QETt_QT5} and \ref{figur:QETt_QT9}
illustrate these measurements
	exemplarily\footnote{The complete set of QETt charts for all templates can be found in a companion document
		in the aforementioned github repository with the four test servers.}
for
	\removable{query}
templates QT5 and QT9,~respectively. 
%
%
As a last experiment, we have increased the scale factor from 100 to~125, and then to~150, and measured \aQETq\
	for five randomly selected
queries per template, as illustrated in
	Figure~\ref{figur:QETq_QT5_db101520} for the five queries of QT5.
\removable{%
	In the following,
we discuss the results of these experiments
	by, first, making some general observations and, afterwards, focusing on each optimization technique~individually.}

\subsubsection{General Observations}
\begin{figure}[t]
\centering
\subfloat[\emph{QETt} for QT5]{\label{figur:QETt_QT5}\includegraphics[width=0.4\linewidth]{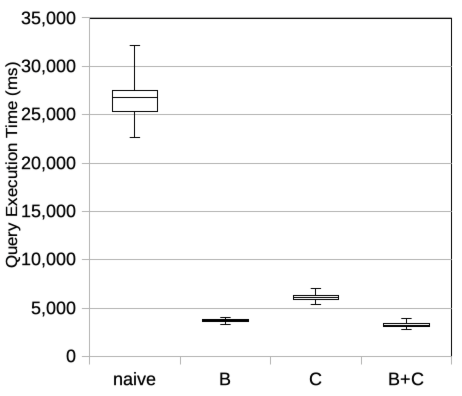}}%
\quad
\subfloat[\emph{QETt} for QT9]{\label{figur:QETt_QT9}\includegraphics[width=0.4\linewidth]{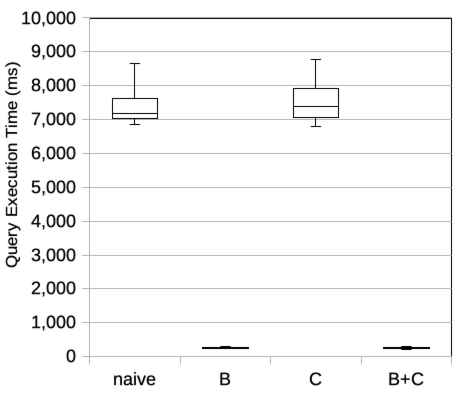}}%
\caption{Comparison of the manually-implemented GraphQL servers in terms of query execution times for 100~queries of the template QT5~(a) and 100 queries of QT9~(b), at scale factor 100.}
\label{Fig:TP_perQT}
\end{figure}

\begin{figure}[b]
	\centering
	\includegraphics[width=\linewidth]{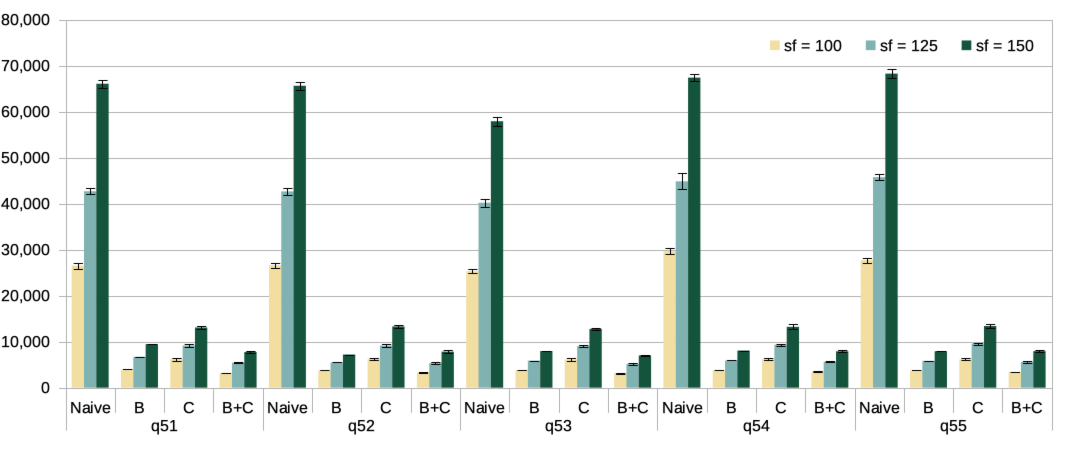}
	\caption{Average query execution times~(\aQETq, in ms) for individual queries of QT5 at increasing scale factors.} 
	\label{figur:QETq_QT5_db101520}
\end{figure}

A first, expected observation is that smaller query execution times result in a greater throughput, as can be seen by comparing the
	\aTPt~(Figure~\ref{Fig:TP_oneClient}) and the corresponding QETt~(Figures~\ref{figur:QETt_QT5}--\ref{figur:QETt_QT9}) that each test server achieves both for QT5 and for QT9.
We also observe that,
for the queries of some query templates, the execution times increase significantly at increasing scale factors%
	~(e.g., QT5, cf.\ Figure~\ref{figur:QETq_QT5_db101520}), whereas for queries of other templates the changes are less substantial.
We explain these differences by the differences in how the respective query result sizes increase at greater scale~factors%
	~(cf.~Section~\ref{sec:sizeOfResult})%
.

\subsubsection{Server-Side Caching}
This optimization aims to reduce the number of requests to the underlying data source by serving repeated requests from a cache. In our experiments we observe a significant benefit of this optimization only for
	the queries of
QT5. The reason why executions of QT5 queries can leverage caching is because the template captures choke point~CP~2.4 \removable{(traversal of relationships that form cycles)}. More precisely, these queries retrieve data about particular graduate students once, and then come back to these graduate students later in a subquery; additionally, for these graduate students, the queries retrieve data about the students' departments, where multiple students belong to the same department. Serv\-er-side caching enables the GraphQL server to avoid requesting the same data for these students and departments multiple times from the database.
\todo{This discussion is not very nice, but it has to do for the moment. If we have more time (and space?), we should discuss the CP~2.5 templates.}



\subsubsection{Server-Side Batching}
The promise of batching is that combining multiple requests to the underlying data source
	reduces the time to perform
these requests. Figure~\ref{Fig:TP_oneClient} shows that
	batching
indeed helps to increase the throughput for queries of
	\removable{templates}
QT2, QT4%
	--%
QT7,~and~QT9.

The choke point that is common to these query templates is~CP~2.1 which captures traversals of 1:N relationship types.
	Hence, the
execution of
	each of these
queries involves at least two resolvers where the first one returns an array of objects and then, for each of these objects, the second resolver is invoked once. If this second resolver performs an SQL request in the context of the given input object, these requests for the different input objects \removable{from the array} can be batched, and that is exactly the case for
	\removable{the queries of}
the aforementioned~templates.

On the other hand, the same is true also for QT12--QT14.
	\removable{The difference,} however, \removable{is that each of} these three templates contains
a filter condition regarding the objects in the corresponding array, and
	the number of objects that satisfy this condition is small.
As a consequence, the resulting number of SQL requests that are batched in these cases is
	also
	small and, thus, batching has~no~effect.
%
\todo{Olaf, another thing to mention here is that naive has big variations in aQETq for different queries whereas batching is more stable -- see Figures~\ref{figur:QETt_QT5}, \ref{figur:QETt_QT9}, and~\ref{Fig:QETq_QRTq_QT5}}

A question that remains is why QT1 is not affected by batching although it covers CP~2.1 as well. The reason is
	that, in this case,
none of the resolvers that are invoked multiple times issues any SQL requests%
	~(the data they use has already been fetched by parent resolvers)%
. Hence, we conclude that batching addresses CP~2.1 for queries in which any subquery that follows a traversal of a 1:N relationship requires further requests to the underlying data source.

\subsection{Evaluation of Connection Pooling} \label{subsec:UC2}
	Our aim with the second use case is
to demonstrate that \LinGBM\ can be employed to evaluate the effectiveness of approaches to achieve read scalability of a GraphQL server. While there is a wide range of options to this end, we consider a simple option
	for the purpose of demonstrating this use case,
namely, the option to
	increase
the number of connections between the GraphQL server to the database~server.

\subsubsection{Experiments}
\begin{figure}[t]
  \centering
  \subfloat[QT3, max DB connections: 10]{\label{figur:tp_QT3}\includegraphics[width=0.4\linewidth]{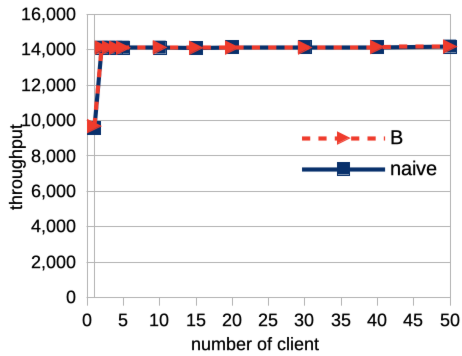}}
\quad
  \subfloat[QT5, max DB connections: 10]{\label{figur:tp_QT5}\includegraphics[width=0.4\linewidth]{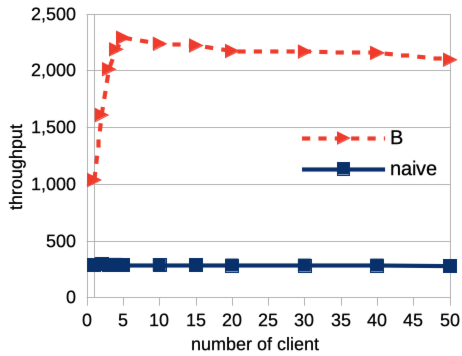}}
\\[2mm]
  \subfloat[QT11, max DB connections: 10]{\label{figur:tp_QT11}\includegraphics[width=0.4\linewidth]{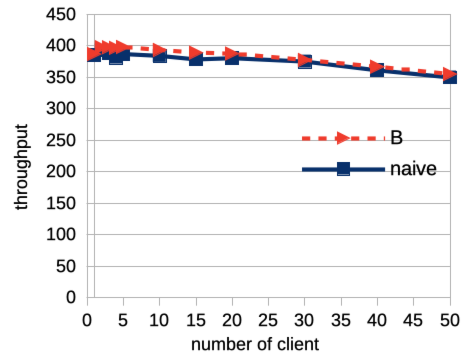}}
\quad
  \subfloat[QT5, server with batching (B)]{\label{Fig:TP_QT5_connectionPooling}\includegraphics[width=0.4\linewidth]{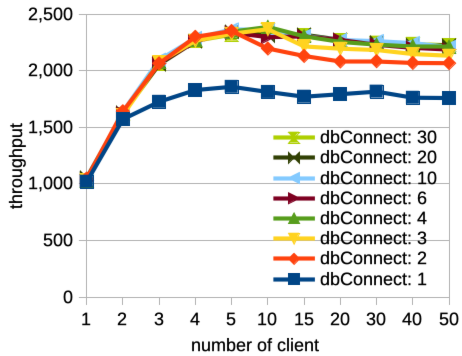}}
  \caption{Average throughput~(\aTPt) achieved for an increasing number of concurrent clients, at scale factor 10.}
  \label{Fig:TP_perQT_diffClients}
\end{figure}

As a first experiment, to understand how
	an increasing
number of clients affects the performance of our man\-u\-al\-ly-im\-ple\-ment\-ed GraphQL servers, we have repeated the first throughput experiment with an increasing number of clients that issue sequences of GraphQL queries concurrently~(%
	first two \removable{clients}, then three, four, five, ten,
15, 20, 30, 40, and 50). 
Figures~\ref{figur:tp_QT3}--\ref{figur:tp_QT11} illustrate these measurements for both the naive server and the server with batching, for the queries of
	templates
QT3, QT5, and QT11, respectively.
Thereafter, for the server with batching, we have repeated this experiment with different values for the maximum number of database connections%
	. Figure~\ref{Fig:TP_QT5_connectionPooling} illustrates these measurements for the queries of%
~QT5~(note that the x-axis in this chart is stretched to better see the measurements for smaller numbers of clients). For these experiments we used a smaller dataset~(%
	scale factor~10)
because, in some cases for the bigger dataset, the test servers became overloaded when serving multiple clients; in particular, this was the case for queries that have much bigger results at greater scale factors~(e.g.,~QT5).

\subsubsection{Observations}
For QT3, and both server variants, we observe that the throughput increases when going from one client to two, but then it does not increase further when adding more clients. We explain this behavior as follows: QT3 queries traverse along three N:1 relationship types and, thus,
	have results that consist of a single leaf node~(cf.\ Table~\ref{tab:estimated-no-of-leaf-nodes}), and batching cannot be leveraged for these queries.
During the execution of each such query, the GraphQL server issues four SQL requests to the database server, one after the other. Hence, with the default connection pool size of~10, the queries of two clients can be served concurrently without any interference. However, when aiming to serve three clients or more, the executions of the concurrent queries are competing for the available database connections and, thus, the throughput stagnates.

For QT5, without batching, the naive test server issues several hundred SQL requests per GraphQL query. In this case, the limited number of available database connections becomes a bottleneck already for one client. In contrast, when using batching, the test server needs only three SQL requests for each QT5 query and, thus, the throughput starts to increase when serving more than one client concurrently. In comparison to QT3, however, for QT5 queries, fetching data from the database is not the only major task of the test server but, instead, the fetched data also needs to be combined into larger result trees. As a consequence, even if concurrent query executions compete for the available database connections, the overall throughput increases up to five clients~(not only up to two as for QT3). However, when increasing the number of concurrent clients beyond five, the throughput starts to drop slightly.
	This is caused by the fact that the batched SQL requests fetch more data, which results in increased waiting times for database connections to become available again. Then, 
constantly switching between requests for different concurrent queries means that the waiting times of each concurrent query execution are affected more and more as the number of concurrent query executions increases.

QT11 is an even more extreme example of this behavior%
	. For
each query of this template, the test servers issue a first SQL request that fetches thousands of objects and, thereafter, each of these objects results in another, separate request~(our current implementation of batching does not cover these because they are
	related to a filter condition).

If we now consider the option to vary the 
	number of available database connections%
~(cf.\ Figure~\ref{Fig:TP_QT5_connectionPooling}), we make two observations in our setting for QT5: First, a connection pool size that is smaller than the default value of~10 causes the throughput to drop already for smaller numbers of clients, which is not unexpected of course. Second, however, increasing the
	number of available database connections
%
	\removable{beyond the default value}
does not help to improve the throughput anymore. At this point, the database server becomes the bottleneck.

\subsection{Evaluation of Generic GraphQL Servers} \label{subsec:UC3}
\begin{figure}[h]
  \centering
  %
  \includegraphics[width=\linewidth]{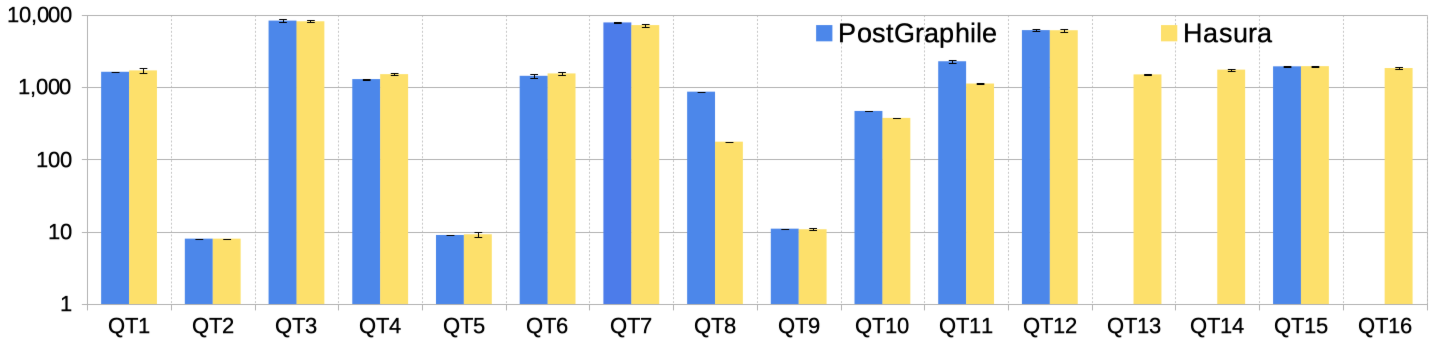}
  \caption{Comparison of two generic GraphQL servers, PostGraphile and Hasura, in terms of average throughput (\aTPt) with one client at scale factor~100 (note that, for QT13, QT14, and QT16, we do not have measurements for PostGraphile because it does not provide the necessary feature for these templates).}
  \label{Fig:G_H_tp_client}
\end{figure}

The third use case is related to Scenario~2 of the benchmark%
	~\removable{(cf.\ Section~\ref{ssec:Design:Scenarios})}%
. We demonstrate this scenario by conducting a preliminary experimental comparison of two generic GraphQL servers%
	, Hasura~(v1.2.2) and PostGraphile~(v4.9.0).
To this end, we set up both of these servers to provide an auto-generated GraphQL API over the \LinGBM\ dataset~($\sfactor\!=\!100$) in the database server used for the previous experiments. Since each of these
	generic GraphQL
servers generates its own GraphQL schema, which is different from our man\-u\-al\-ly-de\-fined \LinGBM\ GraphQL schema, we had to create specific versions of the \LinGBM\ query templates for these two generated schemas. That is, each template had to be rewritten to be expressed in terms of the respective serv\-er-spe\-cif\-ic \hidden{GraphQL }schema such that the resulting queries cover the same choke point(s) and retrieve the same data. While, in the case of Hasura, this was possible for all 16~templates, for PostGraphile we could not translate templates QT13, QT14, and~QT16
	because of
limitations in the generated schema~(in fact,
	to support
QT12
	in PostGraphile we had to use the additional ``con\-nec\-tion-fil\-ter''~plug-in of PostGraphile).
%
To perform the experiment, we have used the same setup as before, but with Docker images\footnote{For Hasura we used the public Docker image. For PostGraphile we created a Docker image based on the instructions on their Website%
	, which was necessary because the publicly available PostGraphile Docker image cannot be used in our setup in which the database server runs in a separate Docker container%
.}
for the two generic GraphQL servers. This way, we have measured the \aTPt\ of both servers for the available templates. Figure~\ref{Fig:G_H_tp_client} illustrates these measurements.

We observe
	that, \removable{in the given setting,} PostGraphile
clearly outperforms Hasura for query templates
	QT8, QT10, and QT11. On the other hand, for QT4,
Hasura is slightly better than PostGraphile, and for the other templates that both of them support%
	~(QT1, QT2, QT6, QT7, QT10, and QT11)%
, none of them is a clear winner%
.
	As a general reason for the observed differences, we point out that the tools use different types of SQL queries internally.

	While this experiment shows that tools such as Hasura and PostGraphile can be evaluated using \LinGBM, we
emphasize that
	we consider a detailed discussion of these observations and, in fact, a more thorough evaluation and comparison of the two tools to be out of scope of this~%
		article.
\todo{Olaf, continue here ... (compare to B+C; however, this makes sense only if we have an explanation for \textbf{QT2---why is B+C so much better here?}) }\bigskip

\section{Concluding Remarks} \label{sec:Conclusions}

This paper introduces \LinGBM, a performance benchmark that captures the key technical challenges~(``choke points'') to be addressed when building an efficient GraphQL server. We have shown statistical properties of the benchmark and demonstrated its applicability for a diverse mix of three different mi\-cro\-bench\-mark\-ing~use~cases.
%
%
%
We emphasize, however, that these are not the only types of use cases for which \LinGBM\ can be employed.

For instance, experiments may be extended to setups in which multiple machines are used~(e.g., to study the effect of remote database servers or to analyze different load-balancing approaches for GraphQL servers).
	Additionally, 
given that the benchmark datasets can also be generated in the form of RDF graphs, approaches to provide GraphQL-based access to such RDF data---and to graph databases in general---can be tested with the benchmark.

\removable{%
Another possible use case is to evaluate and compare different approaches to implement schema delegation~(cf.\ Section~\ref{sssec:Background:Approaches:Delegation}). To this end, different variations of a ``delegating GraphQL server'' may be tested that implement the \LinGBM\ GraphQL schema by delegating requests to
	a generic GraphQL server
that provides
	access to the \LinGBM\ datasets.
}

Further use cases may focus on stress testing of systems by using
	mixed workloads from multiple selected
templates~(%
	for instance,
all templates that cover a particular
	choke~point).
In fact, the definition of
	mixed workloads may go beyond considering only an equal\removable{ number} and uniform distribution of queries from different templates~(as done in Section~\ref{sssec:UC1:MacroLevel}).
\LinGBM\ provides everything needed to design and run experiments with
	workloads that contain
a specific mix of queries that is typical in a particular application scenario~(where some types of queries are more frequent than others). One of our future work tasks is to define and evaluate such workloads for selected application scenarios. Another
	\removable{related}
task will be to extend the benchmark with update operations and possible read-write~workloads.

\begin{acks}
This work was funded by the Swedish Research Council (%
	Vetens\-kaps\-r{\aa}det,
project reg.\ no.~2019-05655), by CUGS~(the National Graduate School in Computer Science, Sweden), and by the CENIIT program at Lin\-k\"oping University~(project no.~17.05).
We also thank Lukas Lindqvist, David {\AA}ngstr\"om, and Markus Larsson for contributing to the development of LinGBM-related software.


\end{acks}

\bibliographystyle{ACM-Reference-Format}
\bibliography{References}

\end{document}